\documentclass[iop,12pt]{emulateapj}
\usepackage[normalem]{ulem}
\usepackage{apjfonts}
\usepackage{epsfig}
\usepackage{natbib}
\usepackage{amsfonts}
\usepackage{amsmath}
\usepackage{multirow}
\usepackage{enumerate}
\usepackage[usenames]{color}
\newcommand\cqg{{Class. Quantum Grav.}}

\def\msun{M$_\odot$}

\def\Dwa{$\,$\uppercase\expandafter{\romannumeral5}$\,$}

\def\sless{\lower2pt\hbox{$\buildrel {\scriptstyle <}
   \over {\scriptstyle\sim}$}}

\def\sgreat{\lower2pt\hbox{$\buildrel {\scriptstyle >}
   \over {\scriptstyle\sim}$}}
\def\sharpnull#1{}

\newcommand{\code}[1]{\texttt{#1}}

\begin{document}
\slugcomment{Draft version \today}

\title{General-Relativistic Simulations of Three-Dimensional Core-Collapse Supernovae}

\author{Christian D. Ott\altaffilmark{1,2,*},
  Ernazar Abdikamalov\altaffilmark{1},
  Philipp M\"osta\altaffilmark{1},
  Roland Haas\altaffilmark{1},
  Steve Drasco\altaffilmark{3,1},
  Evan P. O'Connor\altaffilmark{4,1},
  Christian Reisswig\altaffilmark{1,+},
  Casey A. Meakin\altaffilmark{5,6},
  and Erik Schnetter\altaffilmark{7,8,9}}
  \altaffiltext{1}{TAPIR, Mailcode 350-17,
  California Institute of Technology, Pasadena, CA 91125, USA, 
  cott@tapir.caltech.edu}
\altaffiltext{2}{Kavli Institute for the Physics and
 Mathematics of the Universe (Kavli IPMU), The University of Tokyo, Kashiwa, Japan}
\altaffiltext{*}{Alfred P. Sloan Research Fellow}
\altaffiltext{+}{NASA Einstein Fellow}
\altaffiltext{3}{Grinnell College, Grinnell, IA, USA}
\altaffiltext{4}{Canadian Institute for Theoretical Astrophysics, University of Toronto, Toronto, ON, Canada}
\altaffiltext{5}{Theoretical Division, Los Alamos National Laboratory, Los Alamos, NM, USA}
\altaffiltext{6}{Steward Observatory, University of Arizona, Tucson, AZ, USA}
\altaffiltext{7}{Perimeter Institute for Theoretical Physics, Waterloo, ON, Canada}
\altaffiltext{8}{Department of Physics, University of Guelph, Guelph, ON, Canada}
\altaffiltext{9}{Center for Computation \& Technology, Louisiana State
  University, Baton Rouge, LA, USA}

\begin{abstract}
We study the three-dimensional (3D) hydrodynamics of the
post-core-bounce phase of the collapse of a $27$-$M_\odot$ star and
pay special attention to the development of the standing accretion
shock instability (SASI) and neutrino-driven
convection. To this end, we perform 3D
  general-relativistic simulations with a 3-species neutrino leakage
  scheme. The leakage scheme captures the essential aspects of
  neutrino cooling, heating, and lepton number exchange as predicted
  by radiation-hydrodynamics simulations.  The $27$-$M_\odot$
progenitor was studied in 2D by B.~M\"uller~et~al.\ (\emph{ApJ}
761:72, 2012), who observed strong growth of the SASI while
neutrino-driven convection was suppressed. In our 3D simulations,
neutrino-driven convection grows from numerical perturbations imposed
by our Cartesian grid. It becomes the dominant instability and leads
to large-scale non-oscillatory deformations of the shock front. These
will result in strongly aspherical explosions without the need for
large-scale SASI shock oscillations. Low$-\ell$-mode SASI oscillations
are present in our models, but saturate at small amplitudes that
decrease with increasing neutrino heating and vigor of
convection. Our results, in agreement with simpler 3D
  Newtonian simulations, suggest that once neutrino-driven convection
  is started, it is likely to become the dominant instability in
  3D\@. Whether it is the primary instability after bounce will
ultimately depend on the physical seed perturbations present in the
cores of massive stars. The gravitational wave signal, which we
extract and analyze for the first time from 3D general-relativistic
models, will serve as an observational probe of the postbounce
dynamics and, in combination with neutrinos, may allow us to determine
the primary hydrodynamic instability.
\end{abstract}


\section{Introduction}

\cite{baade:34a} inaugurated core-collapse supernova theory with their
seminal prediction that ``a super-nova represents the transition of an
ordinary star into a neutron star.'' The very basics of this theory,
summarized authoritatively by \cite{bethe:90}, were confirmed by the
observation of neutrinos from SN 1987A \citep{hirata:87,bionta:87}:
The electron-degenerate core of a massive star (with mass $M\sim 8 -
130\,M_\odot$ at zero-age main sequence [ZAMS]), once having reached
its effective Chandrasekhar mass, becomes radially unstable.  Collapse
ensues and, once fully dynamic, separates the core into the
homologous, subsonically contracting inner core and the outer core,
which is supersonically infalling. When the inner core reaches nuclear
density, the nuclear force, which is repulsive at short distances,
leads to a stiffening of the nuclear equation of state (EOS\@).  The
dramatically increased pressure support stabilizes the inner core,
which overshoots its new equilibrium, then rebounds into the still
infalling outer core. This \emph{core bounce} launches a hydrodynamic
shock wave, which, endowed with the kinetic energy of the inner core,
plows into the outer core. Its progression is, however, soon muffled
by energy losses to the dissociation of heavy nuclei and to electron
capture neutrinos that are created and stream out from now
optically-thin regions behind the shock. The hydrodynamic shock thus
succumbs to the extreme ram pressure of the outer core and turns 
into a stalled accretion shock. In the commonly accepted picture of
the \emph{neutrino mechanism} \citep{wilson:85,bethewilson:85,
  bethe:90,janka:07}, the shock is revived by the deposition of
neutrino energy in a layer of net neutrino heating (the \emph{gain}
layer) below the shock. In an alternative scenario, requiring
very rapid progenitor rotation and efficient magnetic field
amplification, a magnetorotational explosion may occur (e.g.,
\citealt{burrows:07b} and references therein). In order to leave
behind a slowly cooling neutron star and not a black hole,
shock revival must occur within a few hundred milliseconds of bounce
\citep{oconnor:11,ugliano:12}.

While the general picture of core-collapse supernova theory may be
well established, details of the explosion mechanism, its
dependence on precollapse conditions and input physics, and its
neutrino and gravitational wave signals\footnote{Both neutrinos
  and gravitational waves may be direct probes of progenitor
  properties, supernova dynamics, and of the explosion mechanism. See,
  e.g.,
  \cite{ott:09,lund:10,lund:12,dasgupta:10b,brandt:11,logue:12,oconnor:13}.}
remain to be determined by detailed first-principles numerical
simulations.

In the case of the neutrino mechanism, modern spherically-symmetric
(1D) simulations with full Boltzmann neutrino transport have shown
that neutrino heating alone fails to drive an explosion in all but the
lowest-mass massive stars
\citep{liebendoerfer:01,rampp:02,thompson:03,sumiyoshi:05,kitaura:06,huedepohl:10,burrows:07c}.
Spherical symmetry, however, is a poor approximation to the situation
after core bounce, even if the initial conditions are nearly
spherically symmetric. The weakening shock leaves behind a negative
entropy gradient, which is expected to lead to convective instability
within milliseconds after bounce (\emph{prompt convection}). Somewhat
later, neutrino heating establishes a negative entropy gradient in the
gain region, leading to \emph{neutrino-driven convection}.  Strong
deleptonization near the neutrinosphere (where the neutrino optical
depth $\tau_\nu \sim 1$; located at the edge of the protoneutron star
at $\sim 10^{11} - 10^{12}\,\mathrm{g\,cm}^{-3}$) establishes a
negative lepton gradient, driving \emph{protoneutron star} convection.

The first full axisymmetric (2D) simulations
\citep{herant:94,bhf:95,janka:95,janka:96,fh:00} showed that 2D
neutrino-driven convection could increase the efficacy of the neutrino
mechanism by increasing the residence time of accreted material in the
region of net neutrino heating and leading to high-entropy turbulent
flow that aids shock expansion.

A new instability, the standing accretion shock instability (SASI),
was discovered by \cite{blondin:03}, who carried out idealized 2D
simulations of an accretion shock using an analytic EOS, neutrino
cooling, but no neutrino heating. In 2D, the SASI leads to large
scale, low-order (in terms of spherical harmonics, $\ell = \{1,2\}$)
deformations of the shock front that vary in time in a predominantly
$\ell = 1$ sloshing-type motion up and down the symmetry axis. These
aspherical motions lead to larger average shock radii, increase the
dwell time of material in the gain region, may lead to secondary
shocks, and are thus generally aiding the explosion mechanism
\citep{ohnishi:06,scheck:06,murphy:08,ott:08,marek:09,mueller:12a}. In 3D,
nonaxisymmetric modes $(m = \{-\ell,...,0,...,\ell\})$ are excited as
well, leading to more complex dynamics and smaller saturation
amplitudes for individual modes \citep{iwakami:08}. In some 3D
simulations, in particular in those that include some initial
rotation, a strong spiral mode ($\ell = 1, m = \pm 1$), capable of
redistributing angular momentum, has been observed
\citep{blondin:07,iwakami:08,iwakami:09,fernandez:10,wongwathanarat:10,rantsiou:11}.

Perturbation theory and carefully controlled numerical experiments
suggest that the SASI is driven by an advective-acoustic cycle in
which entropy and vorticity perturbations are advected from the shock
front to the edge of the protoneutron star. There they trigger the
emission of acoustic perturbations that travel upstream in the
subsonic flow of the postshock region and amplify perturbations in the
shock front, thus creating a feedback cycle that injects power
preferentially into low-order modes (see
\citealt{foglizzo:02,foglizzo:06,foglizzo:07,ohnishi:06,yamasaki:07,
  fernandez:09a,fernandez:09b,scheck:08,guilet:12} and references
therein).  The saturation of the SASI has been proposed to occur via
parasitic Rayleigh-Taylor and/or Kelvin-Helmholtz instabilities that
operate on the entropy gradients and vorticity generated by the SASI
\citep{guilet:10}.

In a real core-collapse supernova, neutrino-driven convection and SASI
overlap in space and may grow at the same time. In the linear regime,
in which seed perturbations are minute, they can be clearly separated:
SASI's fastest growing mode is $\ell = 1$, while convective eddies
will grow with horizontal wavelengths a few times of the entropy scale
height, giving $\ell \sim 7-8$ in the postbounce supernova context
(based on estimates of \citealt{foglizzo:06,herant:92}; see also
\citealt{chandra:61}). In convection, however, all modes are unstable
and will eventually grow to nonlinear amplitudes if convection is able
to develop at all.

How convection and SASI interact in the nonlinear regime, which of
them becomes the dominant instability, how this may depend on the
dimensionality (2D vs.\ 3D), and the ramifications of all this for the
explosion mechanism are open questions that are currently under much
debate. 

\cite{foglizzo:06} argued, based on linear theory, that in the absence
of large (i.e., nonlinear) perturbations the development of
neutrino-driven convection may be suppressed if slowly developing
eddies are advected out of the convectively unstable region before
they can grow significantly. In this scenario, SASI would be the
primary instability. This was also found in the idealized simulations
of \cite{ohnishi:06}, who studied the 2D evolution of an artificially
set up accretion shock with a constant accretion rate and analytic
neutrino cooling, heating, and deleptonization
functions. \cite{scheck:08} performed 2D energy-averaged (gray)
neutrino radiation-hydrodynamics postbounce simulations of a
$15$-$M_\odot$ progenitor star in a carefully controlled setting to
study the development of the SASI\@.  They, too, confirmed the result of
\cite{foglizzo:06} and showed that if sufficiently large ($\gtrsim
1\%$) perturbations from sphericity are present in the upstream flow,
neutrino-driven convection becomes the primary and dominant
instability.

If linearly-growing convection is suppressed by high advection
velocities in the gain region, then one would expect a dependence of
the relative importance of SASI and convection on the postbounce
accretion rate and, hence, on the progenitor star. This was
convincingly confirmed by the recent work of \cite{mueller:12b}, who
carried out full first-principles 2D general-relativistic (GR)
multi-energy radiation-hydrodynamics postbounce simulations of a
$8.1$-$M_\odot$ low-metallicity star with a small core and low
postbounce accretion rate and of a $27$-$M_\odot$ star of solar
metallicity with a large core and high accretion rate. In agreement
with the prediction of \cite{foglizzo:06}, they found strong
convection and absent SASI in the $8.1$-$M_\odot$ star and strong SASI
and nearly absent convection in the $27$-$M_\odot$ progenitor. In both
cases, explosions developed within $\sim$$200\,\mathrm{ms}$ of bounce.

In a different line of research targeted at understanding the
dependence of the neutrino mechanism on dimensionality,
\cite{nordhaus:10} carried out 1D, 2D, and 3D Newtonian collapse
simulations of a $15$-$M_\odot$ progenitor.  They used the simple
analytic heating and cooling prescription introduced by
\cite{murphy:08} (hereafter the MB08 ``light-bulb'' scheme) on the
basis of the work of \cite{janka:01}.  Their 3D simulations did not
show a dominant $\ell = 1$ oscillatory SASI mode observed in 2D
\citep{scheck:06,ohnishi:06,murphy:08}.  Using the critical luminosity
vs.\ accretion rate approach of \cite{burrows:93}, they reported that
in 3D explosions could be obtained at $\sim$$15-25\%$ and
$\sim$$40-50\%$ lower neutrino luminosities than in 2D and 1D,
respectively.

The \cite{nordhaus:10} 3D vs.\ 2D result was not confirmed by
\cite{hanke:12}.  These authors performed Newtonian simulations with
neutrino approximations very similar to the MB08 light bulb, but used a
different 3D hydrodynamics code.  They did not find clear evidence
that 3D effects facilitate the development of an explosion to a
greater degree than the non-radial motions due to SASI and convection
in 2D\@. However, in agreement with \cite{nordhaus:10}, they did not
find large-scale oscillatory low-order modes in their 3D
simulations. They hypothesized that this may be less of a 3D effect
than an effect of the rather simple treatment of neutrino heating and
cooling by \cite{nordhaus:10}. The arguably greatest limitation of the
MB08 light-bulb scheme is its inability to track the contraction of the
protoneutron star, leading to too low advection velocities in the gain
region, thus artificially favoring neutrino-driven convection over the
SASI\@. The results of \cite{takiwaki:12}, whose Newtonian 3D
simulations used a multi-energy approximate neutrino transport scheme,
appear supportive of this assertion. However, these simulations were
carried out with very low resolution and the low-order modes appear to
be clearly oscillatory only at early times.

Using the same MB08 light-bulb approximation for neutrinos and an
updated version of the \cite{nordhaus:10} code, \cite{burrows:12},
\cite{murphy:12}, and \cite{dolence:12} performed and analyzed another
set of 2D and 3D simulations to investigate the roles of SASI and
convection in the postbounce evolution of a $15$-$M_\odot$
progenitor. Comparing 2D and 3D results for the evolution of low-order
fluid mode amplitudes, \cite{burrows:12} showed that at the same MB08
driving luminosity, oscillatory mode amplitudes are much smaller in 3D
than in 2D\@. In models that develop an explosion a non-oscillatory
$\ell = 1$ dipole asphericity grows already in the early postbounce
evolution. Furthermore, they showed that the oscillatory $\ell = 1$
modes observed in 2D -- and generally associated with the SASI --
occur even in the case of a high light-bulb driving luminosity, in
which neutrino-driven convection is the dominant instability. They
argued that in successful explosions by the neutrino mechanism,
neutrino-driven convection should be the dominant instability.
However, for the reasons put forth by \cite{hanke:12} and
\cite{mueller:12b} and discussed in the above, the predictive power of
these light-bulb simulations may be limited.

Ultimately, high-resolution 3D energy-dependent GR neutrino
radiation-hydrodynamics simulations will be needed for final answers
regarding the explosion mechanisms and the role of the various
instabilities involved. Such simulations are computationally extremely
challenging and current attempts are forced to use low spatial
resolution \citep{takiwaki:12,kuroda:12}, the gray approximation
\citep{wongwathanarat:10,mueller:e12,kuroda:12}, and/or employ an
artificial inner boundary, cutting out the protoneutron star core
\citep{wongwathanarat:10,mueller:e12}.

In this paper, we present results from 3D hydrodynamic postbounce
supernova calculations that attempt to strike a balance between the
computationally cheap, but possibly too simplistic light-bulb
approximation and true 3D radiation-hydrodynamics simulations, which
cannot yet be performed without at least partially debilitating
limitations. Our simulations use the \code{Zelmani} core collapse
simulation package \citep{ott:12a} and are fully general relativistic.
We make no symmetry assumptions and use no artificial inner boundary.
We employ a novel computational setup with a multi-block approach that
provides curvilinear grid blocks to track the collapse of the outer
core and Cartesian adaptive-mesh refinement (AMR) grids covering the
central region, including the protoneutron star and the entire
shock. We treat neutrinos in the postbounce phase with an
energy-averaged three-species neutrino leakage scheme with neutrino
heating. The only free parameter of this scheme is a scaling factor in
the charged-current energy deposition rate. As we shall demonstrate,
the leakage scheme captures the essential aspects of neutrino cooling,
neutrino heating, and lepton number exchange. 

We apply \code{Zelmani} to the collapse and postbounce evolution of
the $27$-$M_\odot$ progenitor star that was considered by
\cite{mueller:12b} and shown to be highly susceptible to the SASI in
their fully self-consistent 2D GR simulations. \cite{mueller:12b} find
a SASI-aided explosion that develops within
$\sim$$150-200\,\mathrm{ms}$ after bounce, making this progenitor
ideal for studying the SASI in computationally expensive
high-resolution 3D simulations. We carry out four simulations of the
$27$-$M_\odot$ progenitor, varying the strength of neutrino
heating. We evolve these four models from the onset of collapse to
$\sim$$150-190\,\mathrm{ms}$ after bounce at an effective angular
resolution of $0.85^\circ$ at a radius of $100\,\mathrm{km}$. The
linear resolution at this radius is $\sim$$1.5\,\mathrm{km}$. The
maximum resolution covering the protoneutron star core is
$\sim$$370\,\mathrm{m}$.

We find that neutrino-driven convection is able to grow from the
numerical seed perturbations imposed by our Cartesian AMR approach.
It becomes the dominant instability in the postbounce dynamics of all
of our models. In the case of strong neutrino heating, convection,
which is initially manifest as small-scale cells of rising hotter and
sinking cooler material, develops into large blobs of high entropy
material. These push out the shock and lead to large-scale
\emph{non-oscillatory} shock deformations. We also observe growth of
oscillatory low-$(\ell,m)$ deformations associated with the
SASI\@. However, these saturate at small amplitudes that decrease
further with increasing strength of neutrino heating and vigor of
convection. The SASI remains sub-dominant at all times in our
simulations.  Our results suggest that if neutrino-driven convection
is able to grow in 3D -- which will generally depend on the postbounce
accretion rate and on the seed perturbations present in the flow
\citep{scheck:08} -- it will dominate the postbounce flow. This is
consistent with the results obtained by \cite{burrows:12} with the
simpler light-bulb approach. We extract the gravitational wave signals
generated by accelerated quadrupole mass motions in our models and
find that the strongest component of the signal comes from the initial
burst of convection, which grows on the negative entropy gradient left
behind by the stalling shock.

This paper is structured as follows. In Section~\ref{sec:methods} we
describe \code{Zelmani} and give details on grid setup, EOS, the
leakage/heating scheme, and the progenitor model. In
Section~\ref{sec:results} we present the results of our
simulations. First, in \S\ref{sec:overall}, we give an overview of the
overall postbounce evolution of our models.  We then discuss in
detail the postbounce configurations resulting from our
leakage/heating scheme (\S\ref{sec:checkleak}), the development of
neutrino-driven convection and SASI (\S\ref{sec:convsasi}), various
criteria for neutrino-driven explosions (\S\ref{sec:criteria}), and
the gravitational wave signals extracted from our models (\S\ref{sec:gws}).
We summarize our findings and conclude in Section~\ref{sec:conclusions}.

\section{Methods and Initial Conditions}
\label{sec:methods}

We carry out our 3D GR simulations with the \code{Zelmani} core
collapse simulation package. \code{Zelmani} is based on the
open-source \code{Einstein
  Toolkit}\footnote{\url{http://www.einsteintoolkit.org}}
(\citealt{et:12}), for numerical
relativity and relativistic computational astrophysics. It builds upon
the \code{Carpet} AMR driver \citep{Schnetter-etal-03b} and the
\code{Llama} multi-block system
\citep{pollney:11,reisswig:13a} within the \code{Cactus
  Computational Toolkit} \citep{goodale:03}.

\subsection{Spacetime Evolution and Hydrodynamics}

We evolve the full Einstein equations without approximations in a
$3+1$ decomposition as a Cauchy initial boundary value problem (see,
e.g., \citealt{baumgarte:10book}), using the conformal-traceless BSSN
formulation \citep{baumgarte:99,shibata:95}. A $1+\log$ slicing
condition \citep{alcubierre:00} controls the evolution of the lapse
function $\alpha$, and a modified $\Gamma$-driver condition
\citep{alcubierre:03a} is used for the evolution of the coordinate
shift vector $\beta^i$. The BSSN equations and the gauge conditions
are implemented in the module {\tt CTGamma} using fourth-order
accurate finite differencing. Implementation details are given in
\cite{pollney:11} and \cite{reisswig:13a}. We
  note that in its present form, our evolution system is limited to 3D
  simulations. An extension to 2D along the lines of \cite{baumgarte:12}
may be possible in future work.

We use a flux-conservative formulation of the GR Euler equations,
implemented in the GR hydrodynamics module \code{GRHydro}, which is
part of the \code{Einstein Toolkit} \citep{et:12}. \code{GRHydro} is
an enhanced derivative of the \code{Whisky} \citep{baiotti:05} and
\code{GR-Astro/MAHC} \citep{font:00} codes. It is based on a
finite-volume high-resolution shock-capturing scheme and works with
general finite-temperature microphysical EOS\@.  We employ the
enhanced piecewise-parabolic method for reconstruction of state
variables at cell interfaces
(\citealt{mccorquodale:11,reisswig:13a}) and subsequently
solve approximate Riemann problems to compute intercell fluxes with
the HLLE solver \citep{HLLE:88}. Details are given in
\cite{reisswig:13a}.

Both spacetime evolution and GR hydrodynamics are discretized in a
semi-discrete fashion and coupled with the Method of Lines
\citep{Hyman-1976-Courant-MOL-report} using a multi-rate Runge-Kutta
integrator \citep{reisswig:13a}, providing fourth-order and
second-order accuracy in time for spacetime and GR hydrodynamics,
respectively. The time step is limited by the speed of light and we use
a constant Courant-Friedrichs-Lewy factor of $0.4$.

\subsection{Multi-Block Infrastructure, Adaptive Mesh Refinement, and
Grid Setup}
\label{sec:grid}

\begin{figure}[t]
\centering
\includegraphics[width=0.9\linewidth]{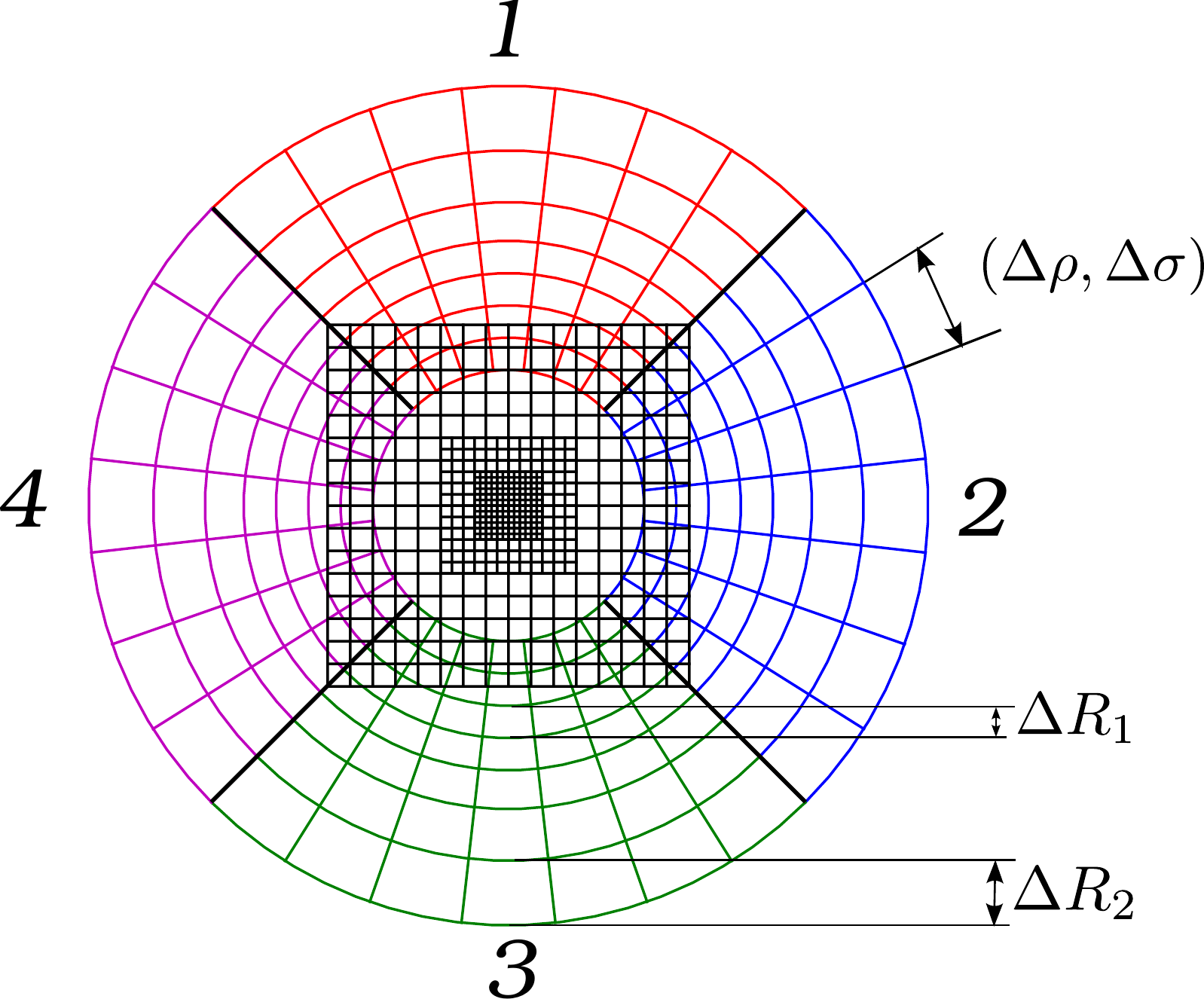}
\caption{Schematic view of a slice through our 3D multiblock grid. Six
  physically curvilinear (four are shown), logically Cartesian
  inflated-cube grids with constant angular, varying radial resolution
  surround a central Cartesian region with five AMR levels (not all
  shown). The third finest level is adjusted to always encompass the
  entire postshock region.}
\label{fig:grid}
\end{figure}

We employ the multi-block infrastructure {\tt Llama}
\citep{pollney:11,reisswig:13a}, which allows us to cover the
computational domain using a set of overlapping curvilinear grid
blocks that are logically Cartesian but physically
curvilinear (so-called ``inflated cubes''), adapted to the overall
spherical topology of the collapse problem. We employ a set of such
curvilinear blocks to track the collapse of the outer core, while the
interior domain containing the protoneutron star, the postshock
region, and the shock itself is covered by an adaptively refined
Cartesian mesh (see Fig.~\ref{fig:grid} for a schematic view).

The spherical inflated-cube multi-block system discretizes one
spherical shell via six angular grid blocks designed such that one
angular coordinate direction always coincides at inter-block
boundaries. This allows us to use efficient fourth-order
one-dimensional interpolation to update ghost zone information between
neighboring blocks \citep{thornburg:04b}.
Furthermore, this particular multi-block system
offers an almost uniform distribution of points across the sphere
(i.e.~without clustering of points at the poles), thus avoiding
distortions and pathologies associated with standard spherical-polar
grids.

The adaptively refined central Cartesian block is based on
cell-centered and flux-conservative mesh-refinement techniques,
provided by the open-source AMR driver {\tt Carpet}
\citep{Schnetter-etal-03b,reisswig:13a}. AMR is implemented
with subcycling in time, following the approach of
\cite{Berger1984}. We make use of \textit{refluxing}, which correctly
adjusts fluxes at mesh refinement boundaries after the AMR restriction
operation \citep{reisswig:13a}. This ensures that mass,
momentum, and energy fluxes are exactly conserved, even in the
presence of strong shocks and other discontinuities.  To update zones
at AMR boundaries and to initialize new grid points after regridding,
we make use of fourth-order prolongation for the spacetime curvature
variables, and second-order essentially non-oscillatory prolongation
for the matter variables.  As detailed in
\cite{reisswig:13a}, spacetime variables are restricted from
fine onto coarse grids using a third-order polynomial, while matter
variables are restricted via cell averaging.







All simulations are carried out with the same general grid setup.  In
the central region, we use five nested Cartesian grids with a factor
of $2$ in resolution between each of them. The finest grid has a
linear cell size $dx = 0.37\,\mathrm{km}$ and extends out to
$17.7\,\mathrm{km}$.  The second finest grid has a cell size of $dx =
0.74\,\mathrm{km}$ and extends to $59\,\mathrm{km}$, while the third
grid has $dx = 1.48\,\mathrm{km}$ and is set up to adaptively track
the shock, ensuring that shock itself and the turbulent flow in the
gain layer behind the shock are always resolved with no worse
resolution than $dx = 1.48\,\mathrm{km}$. For a shock radius of
$100\,\mathrm{km}$, this corresponds to an effective angular
resolution $dx/R$ of $\sim$$0.85^\circ$.  There are two additional
coarser grids with $dx = 2.95\,\mathrm{km}$ and $dx =
5.9\,\mathrm{km}$ in the Cartesian region, which extends to
$532\,\mathrm{km}$, where it overlaps with the outer spherical cube
grid. The latter's radial cell size $dr$ at its inner boundary is the
same as the $dx$ of the coarsest Cartesian grid it overlaps with. $dr$
is held constant out to a radius of $\sim$$3000\,\mathrm{km}$ and then
smoothly reduced to $dr = 189\,\mathrm{km}$ at the outer boundary at
$\sim$$15000\,\mathrm{km}$.  Each of the six cubed-sphere blocks has
$31$ angular zones each in angle $\sigma$ and $\rho$. This corresponds
to an effective cell size of $\sim$$2.9^\circ$.

We start our simulations at the onset of collapse with only the
coarsest of the Cartesian AMR grids active and progressively activate
the finer grids when the central density in the collapsing core
reaches $3.2\times10^{11}\,\mathrm{g}\,\mathrm{cm}^{-3}$,
$1.3\times10^{12}\,\mathrm{g}\,\mathrm{cm}^{-3}$,
$5.1\times10^{12}\,\mathrm{g}\,\mathrm{cm}^{-3}$, and
$2.0\times10^{13}\,\mathrm{g}\,\mathrm{cm}^{-3}$, respectively.

\subsection{Equation of State}

We employ a tabulated version of the finite-temperature nuclear EOS by
\cite{lseos:91}.  This EOS is based on the compressible liquid-drop
model with a nuclear symmetry energy of $29.3\,\mathrm{MeV}$. We use its
variant with a nuclear compression modulus $K_0$ of
$220\,\mathrm{MeV}$, since it yields a cold neutron star mass-radius
relationship in agreement with current observational and theoretical
constraints (e.g., \citealt{demorest:10,hebeler:10,steiner:10}).

We employ the Lattimer-Swesty EOS at densities above
$10^8\,\mathrm{g\,cm}^{-3}$, where $T\gtrsim 0.5\,\mathrm{MeV}$ at all
times in the core collapse context and nuclear statistical equilibrium
(NSE) holds. At lower densities, we employ the Timmes EOS
\citep{timmes:99} and assume that the matter is an ideal gas composed
of electrons, positrons, photons, neutrons, protons, alpha particles,
and heavy nuclei with the average $A$ and $Z$ given by the
Lattimer-Swesty EOS at the transition density. This is an
approximation and may lead to slightly incorrect pressures in non-NSE
regions that result in changes in the collapse times for the silicon
and carbon/oxygen shells. Ideally, a fully consistent treatment with
multiple advected chemical species, a nuclear reaction network and
transition in and out of NSE with a NSE network as proposed by
\cite{buras:06a} should be implemented. This, however, is beyond the
scope of the present study.

Details on the EOS table and on the implementation of the contribution
of electrons, positrons, and photons, as well as other details of the
construction of the table are described in \cite{oconnor:10}. The
table itself as well as table generation and interpolation routines
are available at \url{http://www.stellarcollapse.org}.

\subsection{Neutrino Treatment}
\label{sec:leakage}


We employ the approximate neutrino treatment
of the open-source code \code{GR1D} \citep{oconnor:10}, which was
adapted to 3D and implemented in the module \code{ZelmaniLeak} by
\cite{ott:12a}. The source code is available from
\url{http://www.stellarcollapse.org}.

\begin{deluxetable*}{cccccccc}
\tablecolumns{5} \tablewidth{0pc} \tablecaption{Key Simulation Parameters and Results} 
\tablehead{
Model & $f_\mathrm{heat}$ & $dx_\mathrm{shock}$ & $d\theta,d\phi$ &$t_\mathrm{end}$&$R_\mathrm{shock,max}$&$R_\mathrm{shock,av}$&$R_\mathrm{shock,min}$\\
      &                  & (km)               & @100\,km            & (ms)  & @$t_\mathrm{end}$ & @$t_\mathrm{end}$ & @$t_\mathrm{end}$\\
      &                  &                    & (degrees)   && (km) & (km) & (km)
}
\startdata
$s27f_\mathrm{heat}1.00\phantom{\_\mathrm{HR}}$  
& 1.00 & 1.48 & 0.85& 184 & \phantom{0}82 & \phantom{0}71 & 62\\
$s27f_\mathrm{heat}1.05\phantom{\_\mathrm{HR}}$  & 1.05 & 1.48 & 0.85& 192 & 259 & 189 & 152\\
$s27f_\mathrm{heat}1.10\phantom{\_\mathrm{HR}}$  & 1.10 & 1.48 & 0.85& 165 & 428 & 306 & 204\\
$s27f_\mathrm{heat}1.15\phantom{\_\mathrm{HR}}$  & 1.15 & 1.48 & 0.85& 154 & 432 & 336 & 267
\enddata
\tablecomments{$f_\mathrm{heat}$ is the scaling factor in the neutrino
  heating rate (Eq.~\ref{eq:heating}), $dx_\mathrm{shock}$ is the
  minimum linear resolution covering the shock and the region interior
  to it, $d\theta,d\phi$ @ $100\,\mathrm{km}$ is the effective angular
  resolution at a radius of $100\,\mathrm{km}$, $t_\mathrm{end}$ is
  the time after core bounce at which the simulation is stopped, and
  $R_\mathrm{shock,max}$, $R_\mathrm{shock,av}$, and
  $R_\mathrm{shock,min}$ are the final maximum, average, and minimum
  shock radius, respectively.}
\label{tab:results}
\end{deluxetable*}

Before core bounce, the primary neutrino emission process is electron
capture on free and bound protons, leading to a reduction of the
electron fraction $Y_e$ in the collapsing core.  We include this
effect and associated changes of the specific entropy in the
approximate way proposed by \cite{liebendoerfer:05fakenu}. He showed,
on the basis of 1D Boltzmann neutrino radiation-hydrodynamics
simulations, that $Y_e$ in the collapse phase can be well
parameterized as a function of rest-mass density $\rho$. This
parameterization shows only small variations with progenitor star and
nuclear EOS\@.  We employ an analytic $Y_e(\rho)$ fit to the results of
1D radiation-hydrodynamics collapse simulations of a $20$-$M_\odot$
solar-metallicity progenitor star of \cite{whw:02} obtained with the
code and microphysics of \cite{buras:06a}. The same $Y_e(\rho)$
profile was used in \cite{ott:07prl,ott:07cqg} and \cite{ott:12a}.

In the late collapse phase, when neutrinos begin to be trapped in the
inner core, and throughout the postbounce phase, momentum exchange
between neutrinos and matter becomes non-negligible.  The effect of
this ``neutrino stress'' is naturally captured by the coupling of
radiation and matter in full neutrino transport calculations (see,
e.g., \citealt{mueller:10}). In our approximate treatment, we must
include it explicitly. We assume that neutrino stress is relevant only
above a fiducial trapping density of $2 \times
10^{12}\,\mathrm{g\,cm}^{-3}$ and approximate the stress as the
gradient of the neutrino Fermi pressure. The stress is then included
as a source term in the GR hydrodynamics equations at each
time-integration substep and the neutrino Fermi pressure is included
in the stress-energy tensor (see \citealt{ott:07cqg} and
\citealt{oconnor:10} for details).

After core bounce, which we define as the time at which the specific
entropy at the edge of the inner core reaches
$3\,k_\mathrm{B}\,\mathrm{baryon}^{-1}$, signaling shock formation, the
simple $Y_e(\rho)$ approximation breaks down and fails to even
qualitatively capture the effects of neutrino processes occurring in
the postbounce phase. Dissociation of iron-group nuclei by the shock
provides a sea of free protons for electrons to capture on, leading to
the neutronization burst of electron neutrinos ($\nu_e$) and a steep
drop of $Y_e$ in the region just outside the nascent protoneutron
star. High temperatures and low $Y_e$ in the lower postshock region
allow for the appearance of positrons that capture on neutrons,
leading to the emission of electron antineutrinos ($\bar{\nu}_e$).
High temperatures in the protoneutron star core lead to 
neutral-current pair emission of neutrinos of all species.

In order to capture the aforementioned processes and their effects in
terms of cooling, heating, and deleptonization in the region behind
the shock, we switch to the neutrino leakage scheme of
\cite{oconnor:10} (based on the work of
  \citealt{rosswog:03} and \citealt{ruffert:96}) at bounce. We
consider three neutrino species, $\nu_e$, $\bar{\nu}_e$, and $\nu_x =
\{\nu_\mu,\bar{\nu}_\mu,\nu_\tau,\bar{\nu}_\tau\}$, where we lump the
heavy-lepton neutrinos together, since they participate only in
neutral current processes and have very similar cross sections in the
core-collapse supernova environment.

The leakage scheme provides approximate energy and number emission and
absorption rates based on local thermodynamics and the optical depth
in the postshock region. Neutrino absorption and emission are ignored
outside the shock. The optical depth requires a non-local calculation,
which we solve in a ray-by-ray way, computing an optical depth
integral $\tau_{\nu_i}$ along radial rays cast into $\theta$ and
$\varphi$ directions (see Fig.~1 of \citealt{ott:12a}) from the
origin.  We then interpolate tri-linearly in $(r,\theta,\varphi)$ to
obtain the optical depth at the centers of Cartesian grid cells.
Ideally, an optical depth calculation should be carried out into all
directions from any given cell and the minimum value should be used as
the optical depth of that cell (see, e.g.,
\citealt{ruffert:96}). However, for situations that are spherical at
zeroth order, like the one considered here, the computationally much
cheaper ray-by-ray approach should be sufficient.  In our simulations,
we employ $37$ rays in $\theta$, covering $[0,\pi]$, and $75$ rays in
$\varphi$, covering $[0,2\pi]$. Each ray has 800 equidistant points to
$\sim 600\,\mathrm{km}$ and $200$ logarithmically spaced points
covering $\sim 600 - 3000\,\mathrm{km}$.

We calculate local free neutrino energy
  ($Q^\mathrm{loc}_{\nu_i}$) and number ($R^\mathrm{loc}_{\nu_i}$)
  emission rates for the capture processes $p+e^- \to \nu_e + n$ and
  $e^+ + n \to \bar{\nu}_e + p$ and the thermal processes $e^- e^+$ pair
  annihilation, plasmon decay, and nucleon-nucleon
  bremsstrahlung. Using the estimate of the optical depth
  $\tau_{\nu_i}$, we compute diffusive emission rates
  $Q^\mathrm{diff}_{\nu_i}$ and $R^\mathrm{diff}_{\nu_i}$ and obtain
  the final energy and number loss predicted by the
  leakage scheme by interpolating
  between free emission and diffusive emission rates,
  \begin{equation}
  \chi^\mathrm{leak}_{\mathrm{eff},\nu_i} =
  \chi^\mathrm{leak}_{\mathrm{loc},\nu_i} / ( 1 +
  \chi^\mathrm{leak}_{\mathrm{loc},\nu_i}/\chi^\mathrm{leak}_{\mathrm{diff},\nu_i})\,\,,
\end{equation} 
where $\chi = Q$ for energy loss and $\chi=R$ for number loss (see
\citealt{rosswog:03b} and \citealt{oconnor:10} for definitions and
details).

We approximately include neutrino heating by charged-current
absorption of $\nu_e$ and $\bar{\nu}_e$ on neutrons and protons,
respectively. For this, we make use of a local heating function
based on the derivations by \cite{janka:01},
\begin{equation}
Q^{\mathrm{heat}}_{\nu_i} = f_\mathrm{heat} \frac{L_{\nu_i}(r)}{4\pi r^2}
S_{\nu} \langle\epsilon^2_{\nu_i}\rangle\, {\rho\over m_n} 
  X_i \left\langle {1 \over
  F_{\nu_i}} \right\rangle e^{-2\tau_{\nu_i}} \,\,.\label{eq:heating}
\end{equation}
Here $L_{\nu_i}$ is the neutrino luminosity incident from below,
$S_{\nu} = 0.25 (1 + 3\alpha^2) \sigma_0 (m_e c^2)^{-2}$, where
$\sigma_0$ is the fiducial weak interaction cross-section $\sim
1.76\times 10^{-44} \,\mathrm{cm}^2$, $\alpha = 1.23$, and $m_e c^2$
is the electron rest mass energy in MeV. $\rho$ is the rest-mass
density, $m_n$ is the neutron mass in grams, and $X_i$ is the neutron
(or proton) mass fraction.  $\langle \epsilon^2_{\nu_i} \rangle$ is
the mean-squared energy of $\nu_i$ neutrinos. We approximate it
by taking the matter temperature $T_{\mathrm{NS},\nu_i}$ at the
$\nu_i$ neutrinosphere (where $\tau_{\nu_i} = 2/3$) and evaluating
\begin{equation}
\langle \epsilon^2_{\nu_i} \rangle = T^2_{\mathrm{NS},\nu_i} \frac{\mathcal{F}_5(\eta_{\nu_i,\mathrm{NS}})}{\mathcal{F}_3(\eta_{\nu_i,\mathrm{NS}})},
\label{eq:erms}
\end{equation}
where the $\mathcal{F}_j$ are Fermi integrals $\mathcal{F}_j(\eta) =
\int_0^\infty dx\, x^j (e^{x-\eta} + 1)^{-1}$, and $\eta_{\nu_i,NS} =
\mu_{\nu_i,\mathrm{NS}}\, (k_\mathrm{B} T_{\mathrm{NS},\nu_i})^{-1}$,
where $ \mu_{\nu_i,\mathrm{NS}}$ is the chemical potential of neutrino
species $\nu_i$ at its neutrino sphere.  The factor $\left\langle
F_{\nu_i}^{-1} \right\rangle$ is the mean inverse flux factor, which
depends on details of the neutrino radiation field. We parameterize it
as a function of optical depth $\tau_{\nu_i}$ based on the
angle-dependent radiation fields of the neutrino transport
calculations of \cite{ott:08} and set $\left\langle F_{\nu_i}^{-1}
\right\rangle = 4.275 \tau_{\nu_i} +1.15$. While the true mean inverse
flux factor will asymptote to $1$ at infinity, this simple fit leads
to values in the postshock region (we include heating only there) in
agreement with \cite{ott:08}. Finally, the factor $e^{-2\tau_{\nu_i}}$
is applied to strongly suppress heating at optical depth above
unity. The leakage scheme implementation in \code{Zelmani} varies
slightly from the original implementation in
\cite{oconnor:10,oconnor:11}. In \cite{oconnor:10,oconnor:11}, leakage
was calculated only inside the shock to avoid unnecessary calculations
outside of the shock where little cooling or heating occurred. To
facilitate easy implementation in \code{Zelmani}, where the
angle-dependent shock radius is evaluated only infrequently, we have
removed the explicit dependence on the shock radius and have replaced
it with a condition on the mass fraction of heavy nuclei: we only
calculate the heating and cooling terms where the heavy nuclei mass
fraction is smaller than 0.5 or the density is higher than
$10^{13}\,\mathrm{g}\,\mathrm{cm}^{-3}$. 

We obtain the neutrinosphere locations and the thermodynamic
conditions for Eq.~(\ref{eq:erms}) from the rays used for the optical
depth calculations. We also solve full leakage problems including
heating along the rays to obtain an estimate for the incident
luminosity $L_{\nu_i}$ needed by Eq.~(\ref{eq:heating}). The estimates
for $L_{\nu_i}$ and $\langle \epsilon^2_{\nu_i} \rangle$ are then
interpolated between rays for the local leakage calculations in
Cartesian grid cells. 

All of the above leakage calculations are carried out operator-split
after the fully coupled spacetime/hydro update and are first order in
time. We find this to be sufficiently accurate and stable, due to the
small time step imposed by the light travel time through the smallest
cell. The energy and lepton number updates are applied to the fluid
rest-frame quantities and we ignore velocity dependence or other
relativistic effects in consideration of the overall very approximate
nature of the leakage scheme. The computationally most expensive
aspect of the leakage scheme is the interpolation of density,
temperature, and electron fraction onto the rays. This interpolation
is executed at every time step in the highly dynamic early postbounce
phase. We later switch to carrying out this interpolation only every
16 fine grid time steps (corresponding to every $\sim 8\times
10^{-6}\,\mathrm{s}$) while continuing to evaluate the local
expressions at every time step.

\subsection{Initial Model}

We simulate core collapse and postbounce evolution in the nonrotating
single-star $27$-$M_\odot$ solar-metallicity model $s27$ of
\cite{whw:02}. We choose this particular model to facilitate
comparisons with the recent 2D results of \cite{mueller:12b}.  As
pointed out by \cite{mueller:12b}, this progenitor has an iron-core
mass\footnote{We define the iron-core mass as the mass coordinate that
  has a $Y_e$ of 0.495} of $\sim$1.5\,$M_\odot$ and a silicon-shell
mass of $\sim$0.18\,$M_\odot$.  According to \cite{oconnor:11}, this
progenitor, having a bounce compactness parameter $\xi_{2.5} = 2.5 (
R[M=2.5\,M_\odot)] / 1000\,\mathrm{km})^{-1} = 0.233$, is a likely
candidate for explosion via the neutrino mechanism.  The recent work
of \cite{ugliano:12} predicts a higher failed core-collapse supernova
rate for the solar metallicity model set of \cite{whw:02} than the
work of \cite{oconnor:11}.  However, they also predict that this
particular presupernova model is a progenitor of a successful
neutrino-driven core-collapse supernova.

Using the spherically-symmetric \code{GR1D} code of
\cite{oconnor:10,oconnor:11} and modifying its leakage scheme to be
identical to what we use in \code{Zelmani}, we find that
$f_\mathrm{heat}=1.18$ is the critical value of the scaling factor in
Eq.~(\ref{eq:heating}) to drive an explosion that sets in at late times
after multiple cycles of radial shock oscillations. $f_\mathrm{heat} =
1.28$ is required to drive an explosion without shock oscillations
that sets in at $\sim$$150\,\mathrm{ms}$ after bounce.

We map model $s27$ to our 3D grid under the assumption that the 1D
profile data represent cell averages and use the radii of cell centers
for interpolation. The initial spacetime is set up under the
assumption of spherical symmetry and weak gravity, using the Newtonian
line element without distinction between areal and isotropic radius.

\section{Results}
\label{sec:results}

\begin{figure}[t]
\centering
\includegraphics[width=0.95\columnwidth]{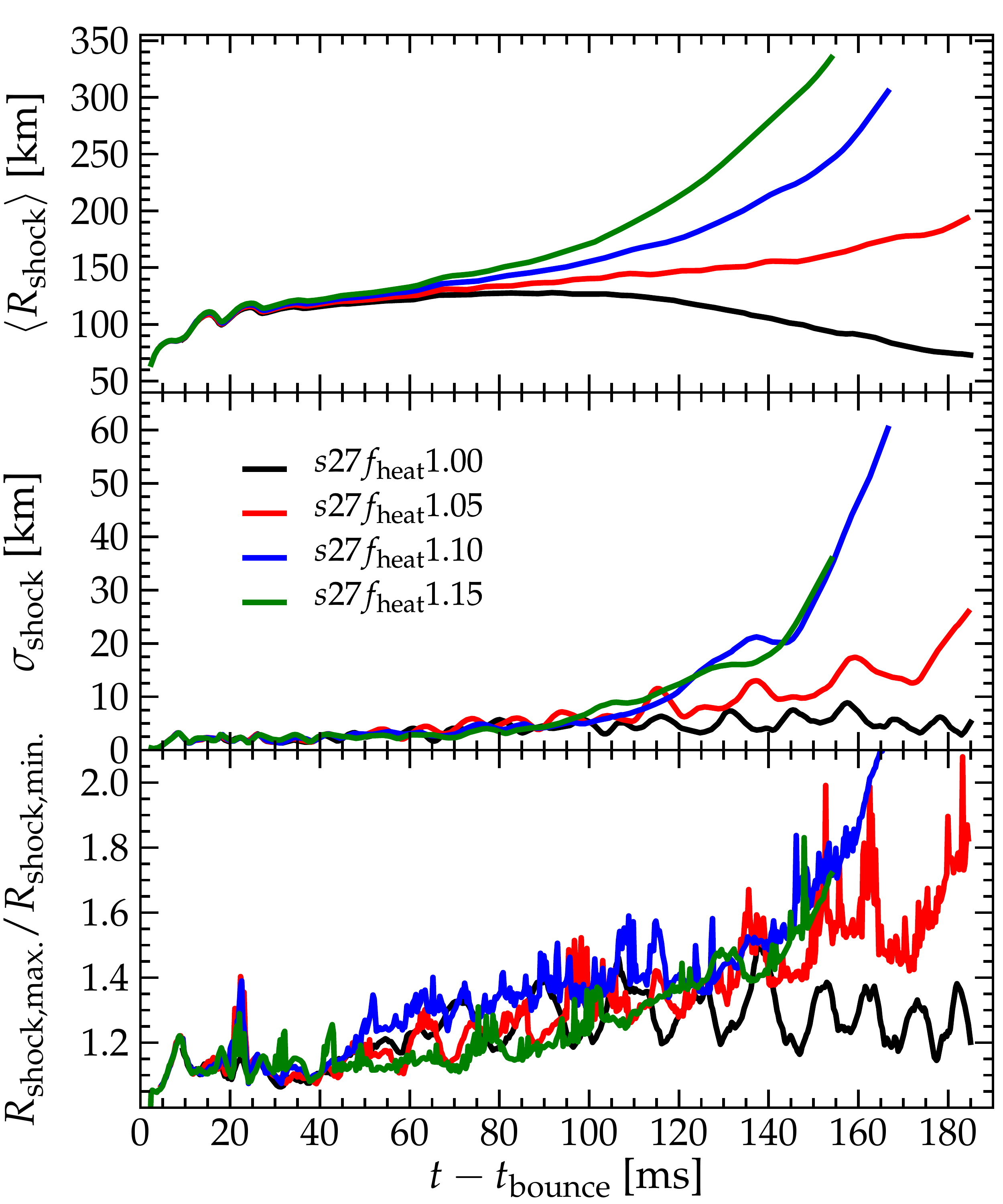}
\caption{Global evolution of the shock in all models. {\bf Top panel}:
  average shock radii $\langle R_\mathrm{shock} \rangle$. {\bf Center
  panel}: standard deviation $\sigma_\mathrm{shock} = \left[(4\pi)^{-1}
    \int d\Omega\, [R_\mathrm{shock} - \langle R_\mathrm{shock}
      \rangle]^2\right]^{1/2}$
  of the shock radii. {\bf Bottom panel}: ratio of maximum
  to minimum shock radius. The shock radii of models with
  $f_\mathrm{heta} \ge 1.05$ exhibit positive trends in their average
  shock radii and have growing $\sigma_\mathrm{shock}$ and ratios
  between maximum and minimum shock radius. Model
  $s27f_\mathrm{heat}1.00$'s shock radius starts decreasing at
  $\sim$$100\,\mathrm{ms}$ after bounce and its
  $\sigma_\mathrm{shock}$ and min/max shock radii ratios oscillate
  around moderate values.}
\label{fig:shockrad}
\vspace{0.5ex}
\end{figure}

\begin{figure}[t]
\centering
\includegraphics[width=0.95\columnwidth]{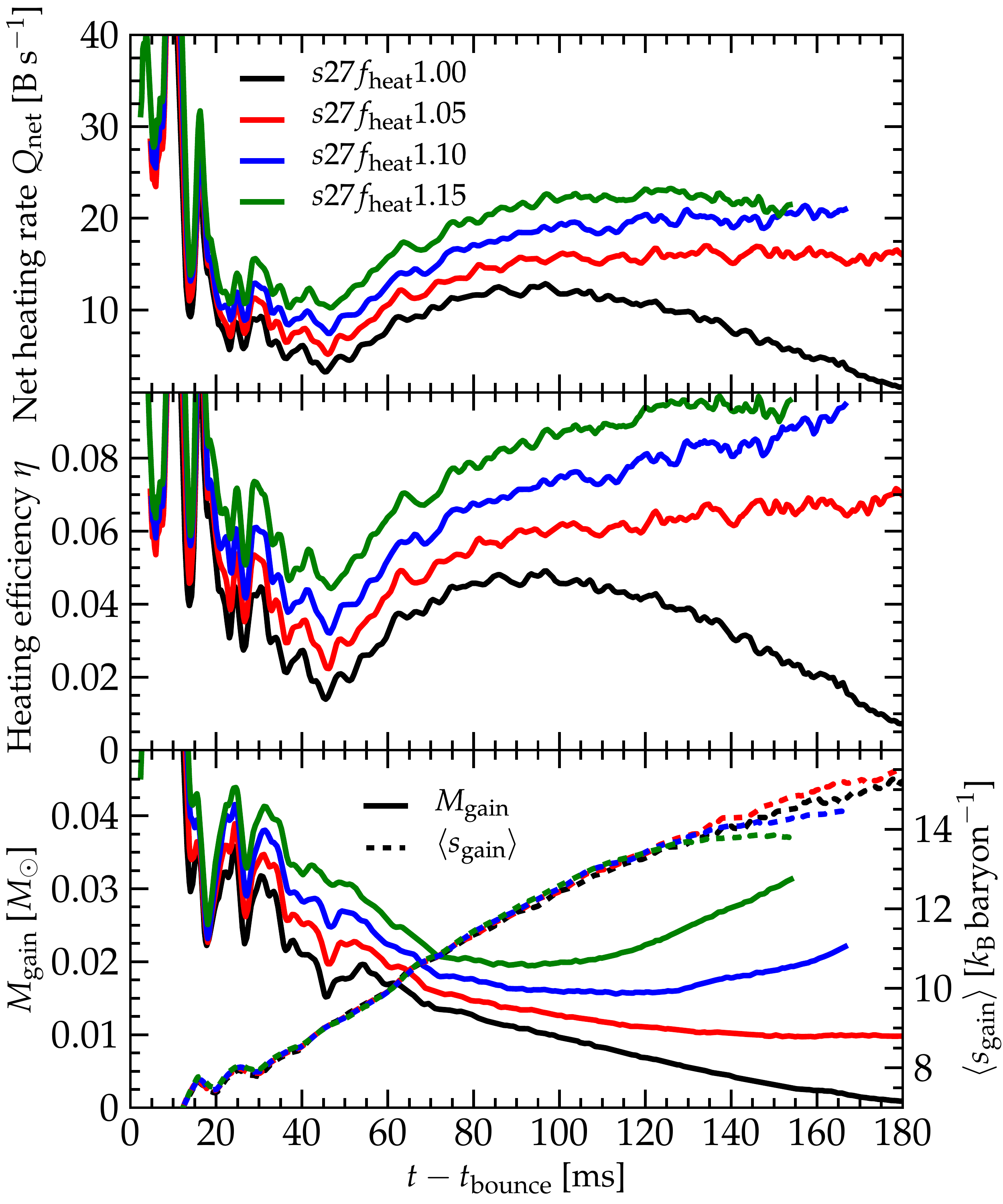}
\caption{Evolution of key integral quantities indicative for the
  strength of neutrino heating. {\bf Top panel}: net neutrino heating
  rate $Q^\mathrm{net}$ (total heating minus total cooling). {\bf
    Center panel}: heating efficiency $\eta$ defined as the net
  heating rate divided by the sum of the $\nu_e$ and $\bar{\nu}_e$
  angle-averaged luminosities incident below the gain layer. {\bf
    Bottom panel}: Mass in the gain layer (left ordinate) and
  density-weighted average specific entropy in the gain layer (right
  ordinate). Heating rate, efficiency, and mass in the gain layer all
  increase monotonically with increasing heating scaling factor
  $f_\mathrm{heat}$. Interestingly, the specific entropy average
  $\langle s_\mathrm{gain} \rangle$ in the gain layer does not exhibit
  such a dependence on $f_\mathrm{heat}$ and the $\langle
  s_\mathrm{gain} \rangle$ curves of all models are nearly identical
  until $\gtrsim 100\,\mathrm{ms}$ after bounce, at which point the
  overall hydrodynamic evolutions have diverged.}
\label{fig:heat}
\vspace{0.5ex}
\end{figure}

\begin{figure*}[t]
\centering
\includegraphics[width=0.95\textwidth]{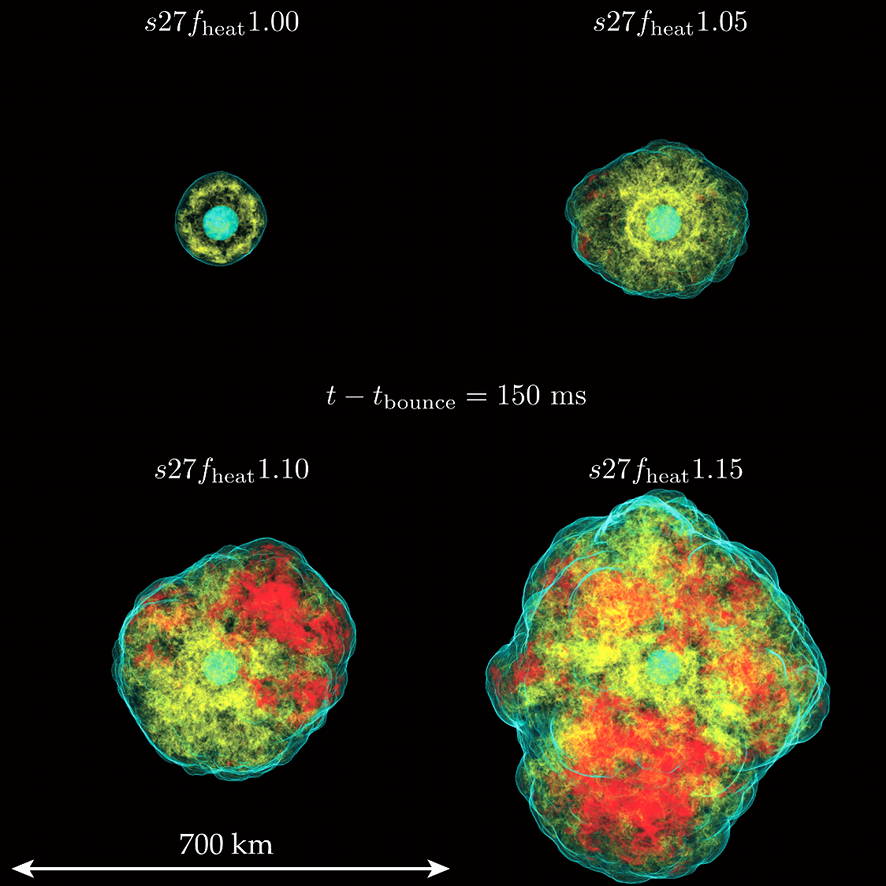}
\caption{3D Volume renderings of the specific entropy at
  $\sim$$150\,\mathrm{ms}$ after bounce in the four simulated
  models. The $z$-axis of the frames is the vertical, $x$ is the
  horizontal and $y$ is into the frame. The scale of the frames is
  $700\,\mathrm{km}$ on a side.  The colormap is chosen such that cyan
  corresponds to a moderate specific entropy of $\sim$$
    4.3\,k_\mathrm{B}\,\mathrm{baryon}^{-1}$, indicating the shock
    front and low-entropy regions near the protoneutron star. Regions
    in yellow indicate higher entropy gas at $s \sim
    16\,k_\mathrm{B}\,\mathrm{baryon}^{-1}$ and red regions correspond
    to gas with $s \sim 20\,k_\mathrm{B}\,\mathrm{baryon}^{-1}$.
    These values are chosen to highlight the surface of the shock and
    gas at a representative ``intermediate'' and a representative
    ``high'' specific entropy.  Note the large scale global
  asymmetries and the many small blob-like protrusions in the shock
  fronts of models whose shock has reached large radii.}
\label{fig:volume}
\end{figure*}

\subsection{Overall Postbounce Evolution}
\label{sec:overall}

We simulate core collapse and bounce of the $27$-$M_\odot$ progenitor
in full 3D with adaptive mesh refinement, adding refinement levels as
the collapse towards a protoneutron star proceeds (see
\S\ref{sec:grid}).  Core bounce, defined as the time when the entropy
at the edge of the inner core reaches
$3\,\mathrm{k_B}\,\mathrm{baryon}^{-1}$, occurs at
$\sim$$299\,\mathrm{ms}$. At bounce, we switch from the $Y_e(\rho)$
parameterization of \cite{liebendoerfer:05fakenu} to the
leakage/heating scheme described in \S\ref{sec:leakage}.  This scheme
includes a scaling factor $f_\mathrm{heat}$ in the neutrino energy
deposition rate (Eq.~\ref{eq:heating}). We carry out four long-term
postbounce simulations, choosing $f_\mathrm{heat} =
\{1.00,1.05,1.10,1.15\}$ to study the influence of changes in the
heating rate on the postbounce evolution. All models are labeled
according to their value of $f_\mathrm{heat}$. For example,
$s27f_\mathrm{heat}1.00$ is the model with $f_\mathrm{heat}=1.00$.
All models are evolved to $\gtrsim 150\,\mathrm{ms}$ after bounce and
for as long as our computer time allocations allow at a cost of $\sim
25,000$ CPU hours per millisecond of physical postbounce time (see
Table~\ref{tab:results}).

\begin{figure*}[t]
\centering
\includegraphics[width=0.8\textwidth]{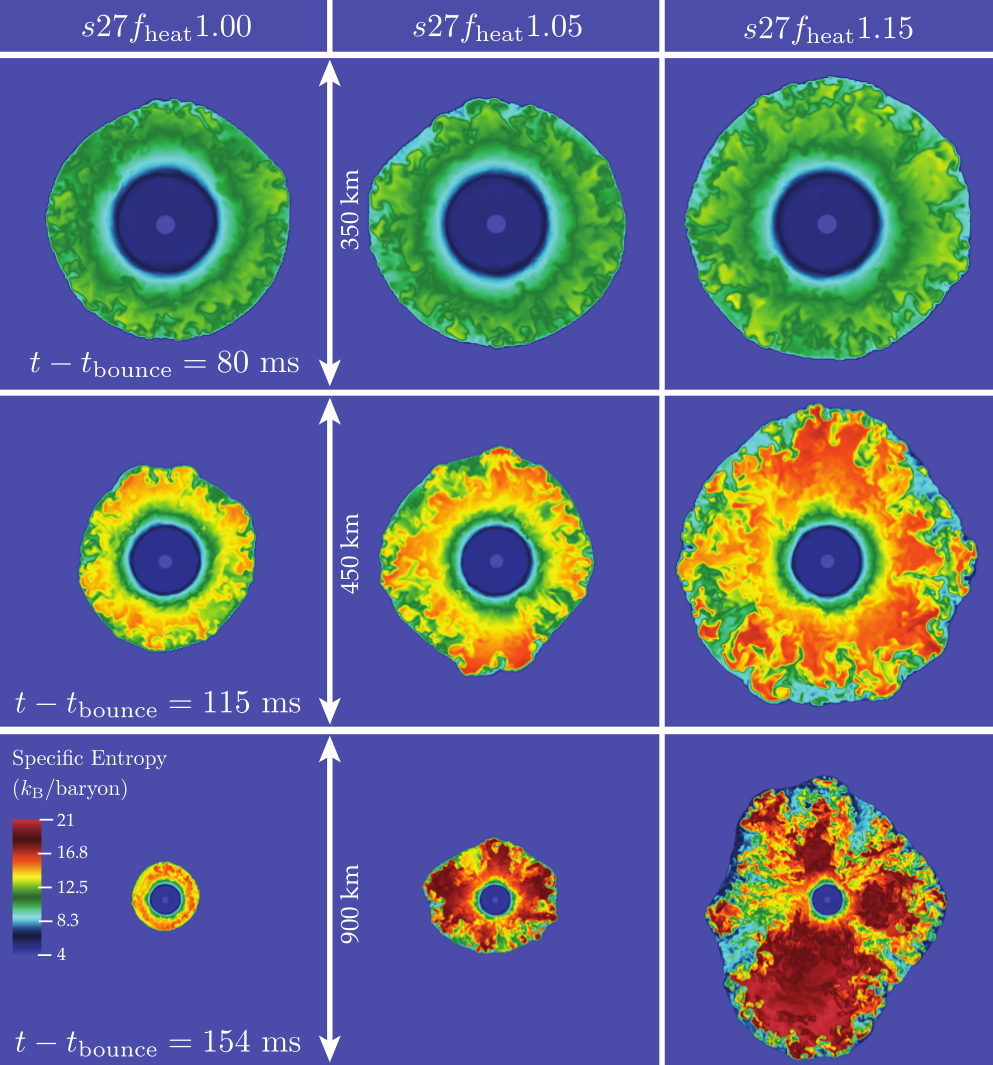}
\caption{Colormaps of the specific entropy in the $x$-$z$ plane in
  models $s27f_\mathrm{heat}1.00$ (left column),
  $s27f_\mathrm{heat}1.05$ (center column), and
  $s27f_\mathrm{heat}1.15$ (right column) at $80$, $115$, and
  $154\,\mathrm{ms}$ after core bounce. The linear scales of the three
  vertical panels are $350$, $450$, and $900$~km at these three times.
  The values of the specific entropy in the convectively unstable gain
  region increase with time in all simulations.  Model
  $s27f_\mathrm{heat}1.00$ exhibits a stagnant shock and only small
  deviations from sphericity. The average shock radius is secularly
  growing in model $s27f_\mathrm{heat}1.05$ with slightly stronger
  neutrino heating and the shock is more aspherical. Model
  $s27f_\mathrm{heat}1.15$ is on track to explosion and exhibits, at
$154\,\mathrm{ms}$ after bounce, a strongly deformed shock with a
single large high-entropy bubble.}
\label{fig:visit_entropy}
\end{figure*}

In the top panel of Fig.~\ref{fig:shockrad}, we show the
angle-averaged shock radius as a function of time in the four
simulated models. After the early dynamic expansion phase, shock
expansion stagnates and the shock stalls at $100 - 130\,\mathrm{km}$
about $40\,\mathrm{ms}$ after bounce. Up to this point, the evolution
is virtually independent of $f_\mathrm{heat}$. In the subsequent
period of quasi-stationary evolution, the gain layer develops and
neutrino heating drives a secular shock expansion.

The neutrino luminosity emitted from the protoneutron star core and
provided by accretion is identical in all models. Hence, as shown in
the top panels of Fig.~\ref{fig:heat}, there is a monotonic increase
with $f_\mathrm{heat}$ in the net neutrino heating rate
$Q_\mathrm{net}$ and in the heating efficiency $\eta = Q_\mathrm{net}
(L_{\nu_e} + L_{\bar{\nu}_e})^{-1}$, where we use the angle-averaged
luminosities at the base of the gain layer.  Varying $f_\mathrm{heat}$
by a moderate $15\%$ from $1.00$ to $1.15$ results in $\sim$$100\%$
more total net heating, since the increase in the local energy
deposition rate results in an expanded gain layer with more mass that
is able to absorb net neutrino energy (cf.\ bottom panel of
Fig.~\ref{fig:heat}).

The quantitative differences in neutrino energy deposition translate
to qualitative differences in the shock evolution. In model
$s27f_\mathrm{heat}1.00$, which has the least heating, shock
stagnation turns into recession and the average shock radius decreases
to $\sim$$70\,\mathrm{km}$ at the end of the simulation. The situation
is very different in models $s27f_\mathrm{heat}1.10$ and
$s27f_\mathrm{heat}1.15$, which both show expanding average shock
radii, surpassing $300\,\mathrm{km}$ at the end of their simulations
and trending towards explosion. Model $s27f_\mathrm{heat}1.05$ is
somewhere in between, but has a slowly, but steadily increasing
average shock radius that reaches $\sim$$190\,\mathrm{km}$ at the
end of the simulation. 

The center and bottom panels of Fig.~\ref{fig:shockrad} display simple
measures of the asphericity of the shock: $\sigma_\mathrm{shock}$, the
angular standard deviation of the shock radius, and
$R_\mathrm{shock,max} / R_\mathrm{shock,min}$, the ratio of maximum to
minimum shock radius. Both quantities show an initial local maximum at
$\sim$$8\,\mathrm{ms}$ after bounce, which is due to an initial
transient large $\ell = 4$ deformation of the shock front caused by
the Cartesian grid employed in our simulations. We will discuss this
further in \S\ref{sec:convsasi}. In the first $40\,\mathrm{ms}$ after
bounce, all models show very similar small deviations of the shock
from spherical symmetry. Differences between models begin to be
apparent at the same time their average shock radii begin to
diverge. Models $s27f_\mathrm{heat}1.10$ and $s27f_\mathrm{heat}1.15$
exhibit very large asymmetries with $\sigma_\mathrm{shock} \sim
30\,\mathrm{km}$ and almost a factor of two in radius between maximum
and minimum shock radius at the end of their simulations (see
Table~\ref{tab:results} for final minimum, maximum, and average shock
radii for all models). Model $s27f_\mathrm{heat}1.05$ also shows
growing asymmetry with increasing postbounce time, similar to the
models with $f_\mathrm{heat} = 1.10$ and $1.15$, but, at least in
$\sigma_\mathrm{shock}$, a periodicity is visible, which is lacking
completely or is occurring at a much smaller level in the two models
with larger $f_\mathrm{heat}$.  In model $s27f_\mathrm{heat}1.00$,
which does not show a positive trend in its shock radius, the
deviations of the shock from sphericity remain small and maximum and
minimum shock radius differ, on average, by $\sim$$20\%$ and this
average difference does not grow until the end of the simulation.
There is, however, clear oscillatory behavior (with a short period of
$\sim$$15\,\mathrm{ms}$) in this model's shock radius variations,
which may be indicative of SASI activity. We shall investigate this
further in \S\ref{sec:convsasi}.

In Fig.~\ref{fig:volume}, we present volume renderings of the specific
entropy at $\sim$$150\,\mathrm{ms}$ after bounce for all four
models. The renderings are all plotted at the same scale to emphasize
the differences in shock radius and 3D geometry between the models.
The color map and rendering opacity are chosen to emphasize ($i$)
regions with specific entropy of
$\sim$$4.3\,k_\mathrm{B}\,\mathrm{baryon}^{-1}$ (cyan), ($ii$) regions
with a representative ``intermediate'' specific entropy of
$\sim$$16\,k_\mathrm{B}\,\mathrm{baryon}^{-1}$ (yellow), and, ($iii$)
regions with a representative ``high'' specific entropy of
$\sim$$20\,k_\mathrm{B}\,\mathrm{baryon}^{-1}$ (red).  Red and yellow
thus mark gas in the high-entropy gain layer, while cyan indicates the
shock front and an iso-entropy surface at the edge of the protoneutron
star. While the shock appears nearly spherical in model
$s27f_\mathrm{heat}1.00$, it is clearly deformed in model
$s27f_\mathrm{heat}1.05$, and strongly so in models
$s27f_\mathrm{heat}1.10$ and $s27f_\mathrm{heat}1.15$. One also notes
that in the latter two models the highest-entropy gas is concentrated
in the region of greatest expansion while it is more evenly spread out
in the other models. The shock deformation in these models is clearly
dominated by low-$\ell$ modes, but there is still much smaller-scale
structure in the form of protrusions caused by rising hot gas bubbles
that push out the shock front at local scales. The
  overall morphology of the expanding shock fronts seen in these
  models is similar to what was found by \cite{dolence:12} in
  exploding 3D Newtonian light-bulb models of a $15$-$M_\odot$
  progenitor, but their shock fronts appear to have less small-scale
  structure than ours (cf.~their Fig.~20).

Figure~\ref{fig:visit_entropy} depicts colormaps of 2D $x-z$ slices of
the specific entropy in models $s27f_\mathrm{heat}1.00$,
$s27f_\mathrm{heat}1.05$, and $s27f_\mathrm{heat}1.15$ at $80$, $115$,
and $154\,\mathrm{ms}$ after bounce. Model $s27f_\mathrm{heat}1.10$ is
not shown, but is overall very similar to model
$s27f_\mathrm{heat}1.15$. The evolution towards large shock radii,
large-scale shock deformation, and peak specific entropies of
$\gtrsim$$20\,k_\mathrm{B}$ is obvious in the slices belonging to
models $s27f_\mathrm{heat}1.05$ and $s27f_\mathrm{heat}1.15$. In the
latter, at $154\,\mathrm{ms}$, one notes a large high entropy area
subtending an angle of $\sim$$30^\circ$ and ranging from the gain
radius out to the shock, which has the overall greatest radii in this
region. This is consistent with the volumetric view of this model at
approximately the same time, shown in Fig.~\ref{fig:volume}.

In the bottom panel of Fig.~\ref{fig:heat}, we plot the
density-weighted average of the specific entropy in the gain layer
$\langle s_\mathrm{gain} \rangle$ (dashed lines; right
ordinate). While the heating rates differ strongly between the models,
their $\langle s_\mathrm{gain} \rangle$ remain very similar until
$\sim$$100\,\mathrm{ms}$ after bounce and $\langle s_\mathrm{gain}
\rangle \sim 12\,k_\mathrm{B}\,\mathrm{baryon}^{-1}$. The top row of
Fig.~\ref{fig:visit_entropy} shows that the peak entropy reached in
the gain layer is very comparable among the three displayed models at
$80\,\mathrm{ms}$ after bounce. At $115\,\mathrm{ms}$ and, in
particular, at $154\,\mathrm{ms}$, the situation is different.  The
models with increasing shock radii and shock deformations develop
large regions with specific entropies in excess of
$20\,k_\mathrm{B}\,\mathrm{baryon}^{-1}$ and large spatial
variations. In model $s27f_\mathrm{heat}1.15$, at $154\,\mathrm{ms}$
after bounce, the expanding deformed shock has already swept up cold
gas that now moves through the gain layer, leading to a decreasing
$\langle s_\mathrm{gain} \rangle$ in this model. This is consistent
with the decrease in $\langle s_\mathrm{gain} \rangle$ seen at the
onset of explosion in the 3D and 2D simulations of \cite{hanke:12} and
\cite{dolence:12}.  At the same postbounce time, the shock in model
$s27f_\mathrm{heat}1.00$ has receded to $\sim$$90\,\mathrm{km}$ and
the distribution of specific entropy behind it is much more uniform
than in the other models. Its average entropy continues to increase
despite the decrease in net heating (cf.~top panel of
Fig.~\ref{fig:heat}). This is due to the combined effect of smaller
shock radii and small deformation of the shock front.  The average
specific entropy in model $s27f_\mathrm{heat}1.05$ also grows, since
its net heating rate continues to stay high while its shock
deformation is still moderate and shock expansion has not yet become
dynamical.

\subsection{Protoneutron Star, Neutrino Emission, and Thermodynamics of the Postshock Region}
\label{sec:checkleak}

\begin{figure}[t]
\centering
\includegraphics[width=0.95\columnwidth]{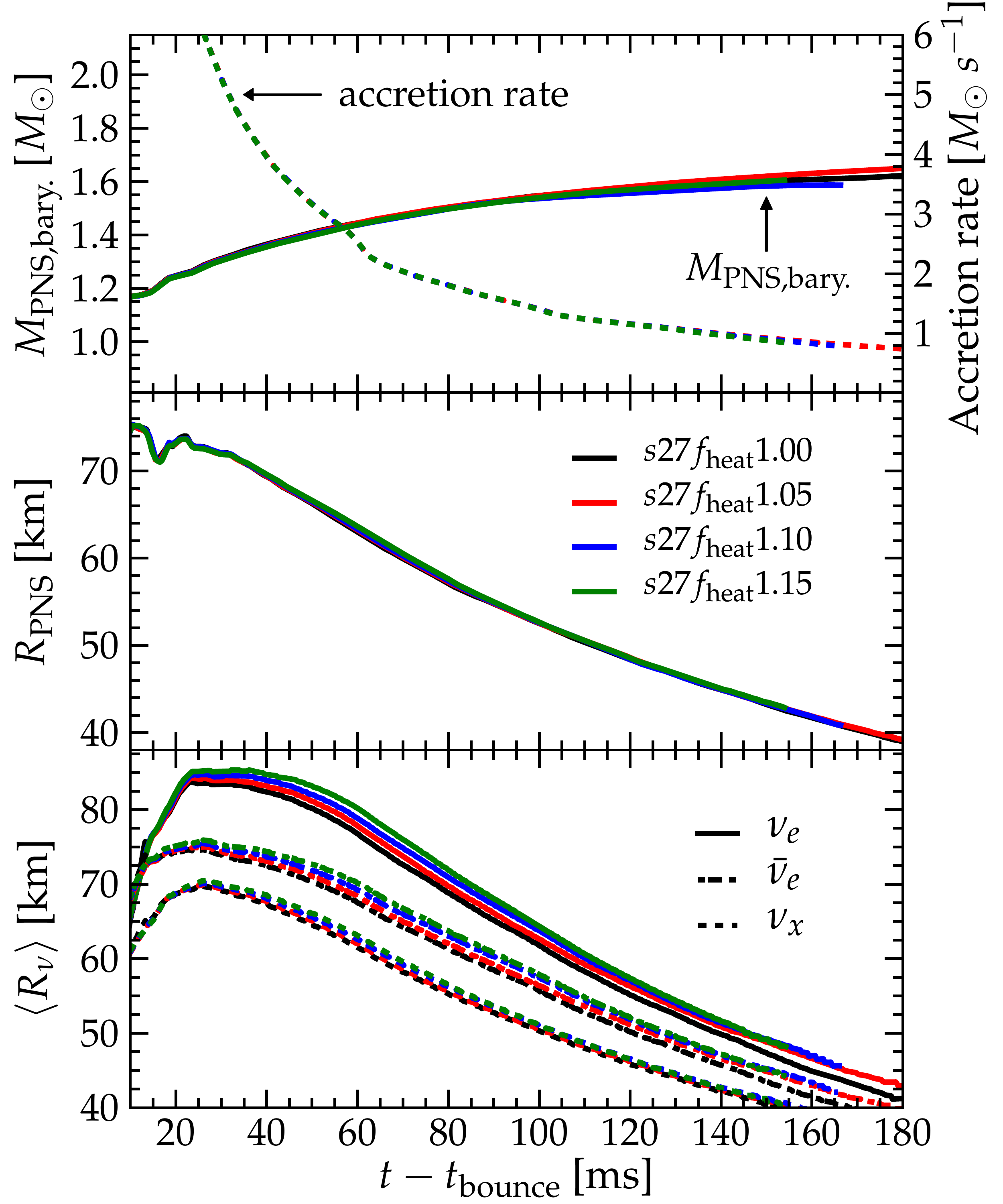}
\caption{{\bf Top panel}: Evolution of the mass accretion rate
  measured at the shock (right ordinate) and baryonic mass of the
  protoneutron star enclosed by the $10^{11}\,\mathrm{g\,cm^{-3}}$
  density isosurface (left ordinate). {\bf Center panel}: Evolution of
  the protoneutron star radius, defined as the location of the
  $10^{11}\,\mathrm{g\,cm}^{-3}$ point on the angle-averaged rest-mass
  density profile. The protoneutron star contracts as neutrino-cooling
  and deleptonizing material is settling on its surface. This is
  expected from 1D and 2D neutrino radiation-hydrodynamics simulations
  (cf.~\citealt{mueller:12a}), but is not captured by the simple MB08
  light-bulb approach (\citealt{richers:13}, \emph{in
      prep.}).  {\bf Bottom panel}: Evolution of the gray,
  angle-averaged neutrinosphere radii $\langle R_\nu \rangle$ as
  predicted by the leakage scheme. The well-known hierarchy $R_{\nu_e}
  > R_{\bar{\nu}_e} > R_{\nu_x}$ is reproduced and the neutrinospheres
  follow the contraction of the protoneutron star as expected from
  full radiation-hydrodynamics simulations (e.g.,
  \citealt{janka:07}).}
\label{fig:pns}
\end{figure}

\begin{figure}[t]
\centering
\includegraphics[width=0.95\columnwidth]{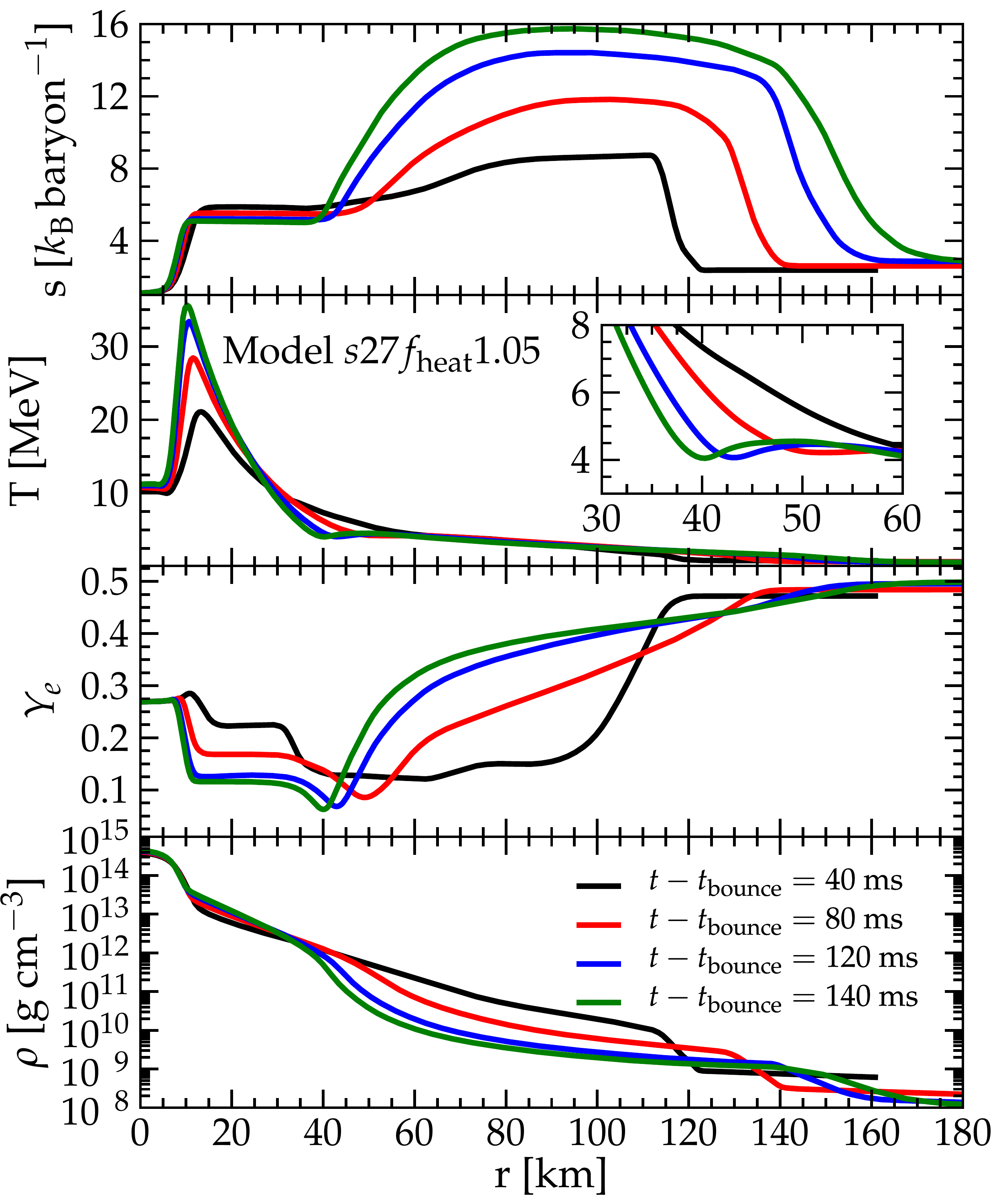}
\caption{Angle-averaged profiles of specific entropy $s$ (top panel),
  temperature $T$ (second panel), electron fraction $Y_e$ (third
  panel) and the rest-mass density $\rho$ (bottom panel) at
  representative postbounce times in model
  $s27f_\mathrm{heat}1.05$. The data are taken from the AMR level
  encompassing the shock and, hence, do not extend to the full
  $180\,\mathrm{km}$ shown at early times. The smoothness of the
  curves is due entirely to the angle averaging.  The profiles show
  the progressive deleptonization and contraction of the outer
  protoneutron star and the development of the high-entropy gain layer
  as is expected from full radiation-hydrodynamics simulations (e.g.,
  \citealt{buras:06a,buras:06b,lentz:12a,mueller:12a}).  Note,
  however, that our leakage/heating scheme tends to somewhat
  overestimate cooling and deleptonization at optical depths of a few,
  leading to a dip in $Y_e$ and a local temperature minimum around
  $40\,\mathrm{km}$. This temperature minimum is shown in a zoomed-in
  inset in the temperature panel.}
\label{fig:profiles}
\end{figure}

The three-species leakage/heating scheme employed in our simulations
goes beyond the MB08 light-bulb approach taken by many recent
3D hydrodynamic studies
(e.g.,~\citealt{nordhaus:10,hanke:12,burrows:12,murphy:12,dolence:12}). These
simulations use analytic cooling functions and neglect important
protoneutron star cooling by $\nu_x$. They also do not take into
account changes of the electron fraction $Y_e$ after bounce
\citep{hanke:12} or do so only via a parameterization of $Y_e(\rho)$,
which cannot account for the strong deleptonization in the region
behind the shock due to electron capture on free protons.  Neutrino
heating is realized in these simulations by an analytic heating
function with spatially and temporally constant neutrino temperature
and luminosity.  An important consequence of these approximations is
that accreted material settling onto the protoneutron star cannot
sufficiently cool, deleptonize and contract
\citep{hanke:12,mueller:12b}. This, in turn, results in too large
shock radii and low advection speeds through the convectively unstable
gain layer that may artificially favor the growth of convection over
SASI \citep{scheck:08,foglizzo:06,mueller:12b}.  Our leakage/heating
scheme is designed specifically to overcome these limitations at
little additional computational cost. We take into account cooling by
$\nu_e$, $\bar{\nu}_e$, and $\nu_x$, account for the change in
electron fraction by $\nu_e$ and $\bar{\nu}_e$ emission and
absorption. Our heating prescription uses the true $\nu_e$ and
$\bar{\nu}_e$ luminosities available at a given position for heating
(as computed by leakage/heating at smaller radii) and the mean-squared
neutrino energies entering the heating rate are determined by
assuming black body emission from the $\nu_e$ and $\bar{\nu}_e$
neutrinospheres, taking the time-changing thermodynamic locations on
these surfaces into account.

While clearly not as sophisticated as recent gray multi-D (e.g.,
\citealt{scheck:08,mueller:e12,kuroda:12}) or energy-dependent (e.g.,
\citealt{ott:08,marek:09,mueller:12a,mueller:12b,takiwaki:12})
neutrino radiation-hydrodynamics calculations, the goal of our
approach is to capture the essential qualitative features correctly
and reproduce quantitative results approximately. In the following, we
investigate the extent to which our scheme lives up to its premise.

In Fig.~\ref{fig:pns}, we plot, for all four models, the time
evolutions of the baryonic mass inside the
$10^{11}\,\mathrm{g}\,\mathrm{cm}^{-3}$ density isosurface (top panel,
left ordinate), the angle-averaged accretion rate measured outside the
shock (top panel, right ordinate), the angle-averaged coordinate
radius of the $10^{11}\,\mathrm{g}\,\mathrm{cm}^{-3}$ density
isosurface (center panel), and the angle-averaged $\nu_e$,
$\bar{\nu}_e$, and $\nu_x$ neutrinosphere radii (where $\tau_{\nu_i} =
1$; bottom panel). The evolutions of protoneutron star mass and
radius, and of the accretion rate are very similar in all models. The
radius if the $10^{11}\,\mathrm{g}\,\mathrm{cm}^{-3}$ isosurface, which we
define as the surface of the protoneutron star following
\cite{mueller:12a}, shrinks from $\sim$$70\,\mathrm{km}$ early after
bounce to $40\,\mathrm{km}$ at $180\,\mathrm{ms}$ after bounce. At the
same time, the enclosed baryonic mass increases from
$\sim$$1.15\,M_\odot$ to $1.55\,M_\odot$. If accretion suddenly
stopped completely at $180\,\mathrm{ms}$, the gravitational mass of
the final, cold neutron star would be $\sim$$1.4\,M_\odot$
\citep{lattimer:01}. The increase in mass and decrease in radius of
the protoneutron star seen in our simulations is qualitatively
consistent with the findings of \cite{mueller:12a} and
\cite{buras:06b} for different progenitors. \cite{mueller:12b}, who
studied the $s27$ progenitor, do not show these quantities. Hence, a
direct quantitative comparison is not possible.

\begin{figure}[t]
\centering
\includegraphics[width=0.95\columnwidth]{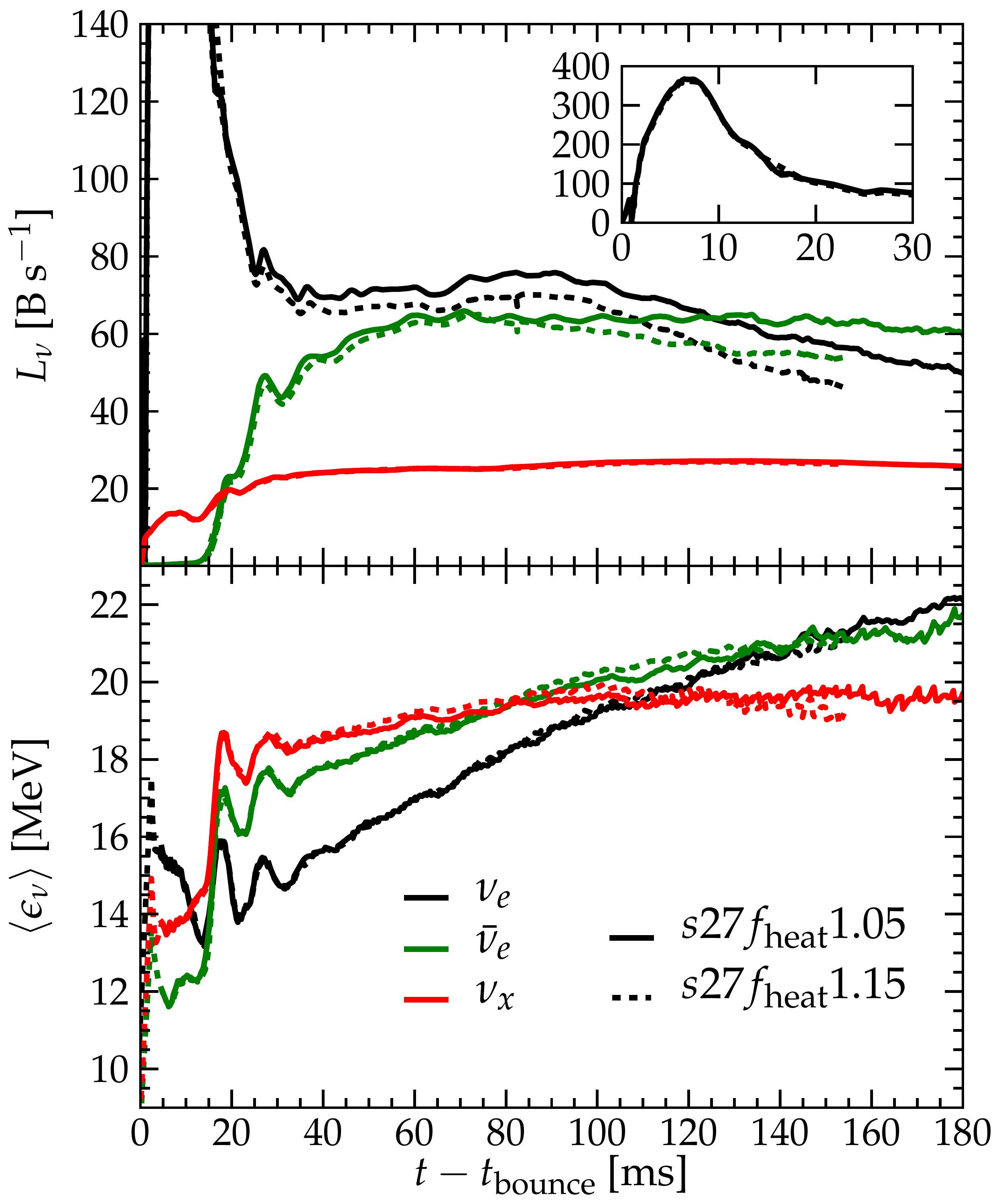}
\caption{{\bf Top panel}: $\nu_e$, $\bar{\nu}_e$, and $\nu_x$
  luminosities as a function of postbounce time in models
  $s27f_\mathrm{heat}1.05$ (solid lines) and $s27f_\mathrm{heat}1.15$
  (dashed lines) as representative examples of our model set. The
  $\nu_e$ and $\bar{\nu}_e$ luminosities in model
  $s27f_\mathrm{heat}1.15$ are somewhat smaller due to the strong
  charged-current absorption in this model. The inset plot shows the
  $\nu_e$ deleptonization peak. Comparing the $L_{\nu_i}$ shown here
  with those provided for the same progenitor in Fig.~8 of
  \cite{mueller:12b} demonstrates that our much more approximate
  neutrino treatment still yields luminosities that agree within
  $\sim$$20\%$ with the results of true radiation-hydrodynamics
  simulations. {\bf Bottom panel}: Evolution of the mean neutrino
  energies $\langle \epsilon_{\nu_i} \rangle$ in the same models,
  obtained via the assumption of black body emission at the respective
  neutrinospheres. After the transient very early postbounce phase,
  the usual hierarchy of mean neutrino energies is established. At
  $\gtrsim$$80-100,\mathrm{ms}$ after bounce, the evolution becomes
  qualitatively incorrect when $\langle \epsilon_{\nu_e} \rangle$ and
  $\langle \epsilon_{\bar{\nu}_i} \rangle$ surpass $\langle
  \epsilon_{\nu_x} \rangle$. Comparison with the results of
  \cite{mueller:12b} shows that this and the overall high predicted
  $\langle \epsilon_{\nu_i} \rangle$ are an artifact of the
  leakage/heating scheme.}
\label{fig:lumeav}
\end{figure}

The angle-averaged neutrinosphere radii given in the lower panel of
Fig.~\ref{fig:pns} show that the leakage scheme correctly reproduces
the well known hierarchy $R_{\nu_e} > R_{\bar{\nu}_e} > R_{\nu_x}$ of
neutrinosphere radii in the postbounce preexplosion phase (e.g.,
\citealt{janka:07}). One notes that models with larger
$f_\mathrm{heat}$ have slightly larger neutrinosphere radii. We
attribute this to their somewhat hotter postshock regions, resulting
in higher opacity.

Model $s27f_\mathrm{heat}1.05$ is intermediate between model
$s27f_\mathrm{heat}1.00$ that fails to explode in the
  simulated time and models $s27f_\mathrm{heat}1.10$ and
$s27f_\mathrm{heat}1.15$, which have rapidly increasing shock radii at
the end of their simulations. We choose $s27f_\mathrm{heat}1.05$ as
our representative model and show, in Fig.~\ref{fig:profiles},
angle-averaged profiles of its specific entropy, temperature, electron
fraction, and rest-mass density at $40$, $80$, $120$, and
$140\,\mathrm{ms}$ after bounce. The smoothness of the profiles is due
entirely to angle averaging. The overall qualitative behavior of all
quantities is as expected from more complex radiation-hydrodynamics
simulations (cf. \citealt{buras:06a} and Fig.~5 of
\citealt{dessart:06pns}). The radial extent and specific entropy of
the gain layer increase with time, while the changes in the specific
entropy below $\sim$$40\,\mathrm{km}$ simply reflect protoneutron star
contraction. The latter is also well captured by the rising
temperature at the protoneutron star edge, indicating compression.
The strong deleptonization of the postshock region caused by the
$\nu_e$ neutronization
burst shortly after bounce is
still visible in the $Y_e$ profile at $40\,\mathrm{ms}$ after bounce.
The outer postshock region re-leptonizes over time due to a slight
dominance of $\nu_e$ over $\bar{\nu}_e$ absorption in the gain
layer. In the lower postshock region ($R \sim 10-45\,\mathrm{km}$),
neutrino cooling and deleptonization continue and, as expected from
more accurate neutrino transport calculations, a strong negative
lepton gradient develops that may drive protoneutron star convection
(e.g., \citealt{dessart:06pns,buras:06b}).

The top panel of
Fig.~\ref{fig:lumeav} shows the total luminosities of $\nu_e$,
$\bar{\nu}_e$, and $\nu_x$ as predicted by our leakage/heating scheme
for models $s27f_\mathrm{heat}1.05$ and $s27f_\mathrm{heat}1.15$,
which we take as representative examples.  Differences between these
models are minor and due to the greater heating in model
$s27f_\mathrm{heat}1.15$. Since \cite{mueller:12b} provide these
luminosities from their 2D simulations in their Fig.~8, we can
directly compare with their results. Their $L_{\nu_e}$ peaks at
$\sim$$385\,\mathrm{B\,s}^{-1}$ ($1\,\mathrm{B} =
10^{51}\,\mathrm{erg}$), while ours peaks at
$\sim$$365\,\mathrm{B\,s}^{-1}$ (a $5\%$ difference). At
$100\,\mathrm{ms}$ after bounce, the \cite{mueller:12b} simulation
suggests $L_{\nu_e} \sim$$62\,\mathrm{B\,s}^{-1}$, while we find
$\sim$$68\,\mathrm{B\,s}^{-1}$ in model $s27f_\mathrm{heat}1.15$
($\sim$$73\,\mathrm{B\,s}^{-1}$ in model $s27f_\mathrm{heat}1.05$), a
$\sim$$10\%$ ($\sim$$20\%$) difference. The $\bar{\nu}_e$ and $\nu_x$
luminosities compare similarly well. The rather good agreement in
total neutrino luminosities with the much more detailed
radiation-hydrodynamics simulation of \cite{mueller:12b} suggests that
our leakage/heating scheme captures the overall neutrino emission and
its energetics in an acceptable way.

The situation is different for the mean neutrino energies $\langle
\epsilon_\nu \rangle$ shown in the lower panel of
Fig.~\ref{fig:lumeav}. We obtain estimates for $\langle \epsilon_\nu
\rangle$ of each species by assuming black body emission from its
neutrinosphere in the same way as for the mean-squared energies that
enter the heating function (Eq.~\ref{eq:heating}). This kind of
estimate is not reliable in the very early, highly dynamical
postbounce phase, but $\sim$$20\,\mathrm{ms}$ after bounce, the usual
hierarchy of neutrino energies $\langle \epsilon_{\nu_x} \rangle >
\langle \epsilon_{\bar{\nu}_e} \rangle > \langle \epsilon_{\nu_e}
\rangle$ is established analogously to the hierarchy of neutrino
sphere radii (cf.~lower panel of Fig.~\ref{fig:pns}). This hierarchy
is, however, broken at times $\gtrsim 80-100\,\mathrm{ms}$, when the
mean $\nu_e$ and $\bar{\nu}_e$ energies exceed the mean energy of
$\nu_x$. This is clearly an artifact of our leakage/heating scheme and
will not happen in nature. It is also not found by \cite{mueller:12b}.
The reason for this incorrect behavior can be understood by
considering the neutrinosphere radii plotted in the lower panel of
Fig.~\ref{fig:pns} and looking at the temperature and $Y_e$ profiles
shown in Fig.~\ref{fig:profiles} for model $s27f_\mathrm{heat}1.05$.
At postbounce times $\gtrsim 80\,\mathrm{ms}$, one notices a global minimum
in $Y_e$ around $40-50\,\mathrm{km}$. At the same location, a local
temperature minimum develops. Both are related and caused by the
inability of the leakage/heating scheme to establish a correct balance
between emission and absorption at optical depths of a
few. Unfortunately, the $\nu_x$ neutrinosphere recedes precisely into
the local temperature minimum, while the $\nu_e$ and $\bar{\nu}_e$
neutrinospheres sit in the local maximum at slightly greater
radii. While the differences in temperature are not large, they are sufficient
to explain the incorrect evolution of the $\langle \epsilon_{\nu_i} \rangle$.

Comparing the values of $\langle \epsilon_{\nu_i} \rangle$ predicted
by the leakage scheme with the results of \cite{mueller:12b} at times
before the qualitative evolution becomes unrealiable, we find that the
leakage scheme systematically overpredicts the mean energies. For
example, at $50\,\mathrm{ms}$ after bounce, in model
$s27f_\mathrm{heat}1.05$, we find $\langle \epsilon_{\nu_e} \rangle
\sim$$16\,\mathrm{MeV}$, $\langle \epsilon_{\bar{\nu}_e} \rangle
\sim$$18\,\mathrm{MeV}$, and $\langle \epsilon_{\nu_x} \rangle
\sim$$18.5\,\mathrm{MeV}$.  At the same time the $\langle
\epsilon_{\nu_i} \rangle$ found by \cite{mueller:12b} are, in the same
order, $9.3\,\mathrm{MeV}$, $12.3\,\mathrm{MeV}$, and
$14\,\mathrm{MeV}$.

In summary, the results shown in this section indicate that the
leakage/heating scheme used in our simulations yields overall
qualitatively correct thermodynamics/stratification in the postshock
region and captures the integral neutrino emission to within
$\sim$$20\%$ of fully self-consistent simulations.  It fails, however,
to yield reliable predictions for the mean neutrino energies, in
particular at later postbounce times. Since energy (and lepton number)
absorption rates depend sensitively on neutrino energy, the
leakage/heating scheme, at least in its present form, cannot be
employed to make reliable predictions of the spectrum of the
emitted neutrinos or the composition of explosion ejecta.

\subsection{SASI and Neutrino-Driven Convection}
\label{sec:convsasi}

\begin{figure*}[t]
\centering
\includegraphics[width=0.92\textwidth]{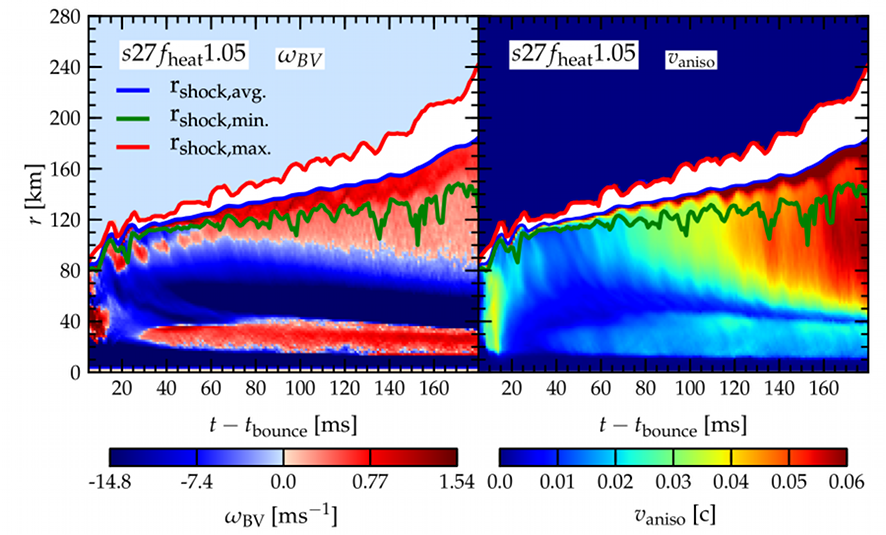}
\includegraphics[width=0.92\textwidth]{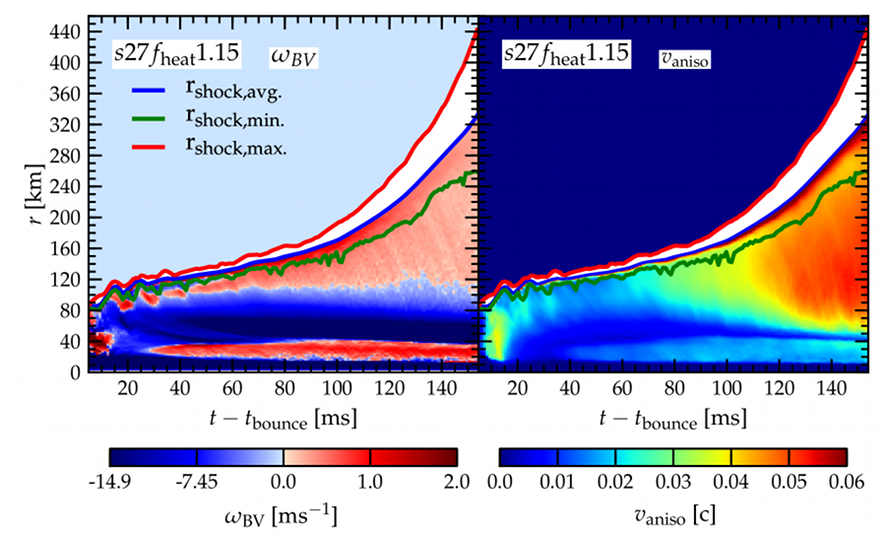}
\caption{Colormaps showing the time evolution of the angle-averaged
  Brunt-V\"ais\"al\"a (BV) frequency $\omega_\mathrm{BV}$ in units of
  $\mathrm{ms}^{-1}$ (Eq.~\ref{eq:omegabv}; left panels) and
  anisotropic velocity $v_\mathrm{aniso}$ in units of the speed of
  light $c$ (Eq.~\ref{eq:vaniso}; right panels) in models
  $s27f_\mathrm{heat}1.05$ (top panels) and $s27f_\mathrm{heat}1.15$
  (bottom panels). Also indicated are the maximum shock radius (red
  curves), the average shock radius (blue curves), and the minimum
  shock radius (green curves). Note the different radial scales of the
  top and bottom panels. We mask out $v_\mathrm{aniso}$ and
  $\omega_\mathrm{BV}$ outside the average shock radius, where they
  are not reliable, since at most angles the radial region is actually
  outside of the shock. Shortly after bounce, the stalling shock
  leaves behind a negative entropy gradient, leading to
  $\omega_\mathrm{BV} > 0$ and thus convective instability, strongest
  at radii between $20$ and $40\,\mathrm{km}$.  Prompt convection
  develops quickly and is strong, as indicated by the large
  $v_\mathrm{aniso}$ in the right panels. Subsequently, convective
  instability and, as shown by the right panels, convection, develops
  in the gain layer and, after $\sim$$30-40\,\mathrm{ms}$ or so, also at
  the edge of the protoneutron star core, due to the negative lepton
  gradient. Note that high $v_\mathrm{aniso}$ at late times prevails
  to significantly smaller radii than the inner radius of convective
  instability, indicating large asymmetries and undershooting of
  decelerating convective plumes.}
\label{fig:bv_vaniso}
\end{figure*}

The recent 2D radiation-hydrodynamics core collapse and postbounce
simulations of the $s27$ progenitor carried out by \cite{mueller:12b}
show a very clear and clean growth of a dominant periodic $\ell = 1$
SASI mode. The relative amplitude (with respect to the average shock
radius) of the $\ell = 1$ mode saturates in their simulations at a
very large $\sim$45\%, indicating a large-scale periodic dipole
deformation. In their simulation, neutrino driven convection is only a
secondary instability that develops in the non-linear phase, but may
be connected with the saturation itself \citep{guilet:10}.

It is now interesting to ask if the SASI is the primary instability
driving asphericity in the $s27$ progenitor also in our 3D simulations
or if neutrino-driven convection dominates early on and possibly
suppresses the growth of coherent SASI oscillations. It is furthermore
interesting to study how the roles and prominence of SASI and
convection depend on the strength of neutrino heating. It is evident
from the discussion in \S\ref{sec:overall} and Figs.~\ref{fig:volume}
and \ref{fig:visit_entropy} that deviations from sphericity develop at
large scales in our models. We shall now take a more quantitative look
at the development of this asphericity.

\subsubsection{Convection}

The local stability of a fluid element to convective overturn is
determined via the Ledoux criterion \citep{ledoux:47},
\begin{equation}
C_\mathrm{L}= - \left(\frac{\partial \rho}{\partial
  P}\right)_{s,Y_l}
\left[
\left(\frac{\partial P}{\partial s}\right)_{\rho,Y_l}
\left(\frac{\mathrm{d} s}{\mathrm{d} r}\right)
+
\left(\frac{\partial P}{\partial Y_l}\right)_{\rho,s}
\left(\frac{\mathrm{d} Y_l}{\mathrm{d} r}\right)
\right],
\label{eq:ledoux}
\end{equation}
which, in the postbounce supernova case, takes into account radial
gradients in specific entropy $s$ and lepton fraction $Y_l = Y_e +
Y_{\nu_e} - Y_{\bar{\nu_e}}$. For simplicity, we set $Y_l := Y_e$,
since our leakage scheme does not keep track of local neutrino
fractions. This approximation may lead to quantitatively incorrect
estimates of $C_\mathrm{L}$ in the protoneutron star where neutrinos
are trapped or partially trapped.  A fluid element is convectively
unstable if $C_{\rm L} > 0$.  The linear growth time for convection
from arbitrarily small perturbations is then given by the
Brunt-V\"ais\"al\"a (BV) frequency,
\begin{equation}
\omega_\mathrm{BV} =
\mathrm{sgn}\left(C_{\mathrm{L}}\right)\sqrt{
\left|\frac{C_{\mathrm{L}}}{\rho}\frac{\mathrm{d}\Phi}{ \mathrm{d} r}
\right|} ,
\label{eq:omegabv}
\end{equation}
where we are following the definition of \cite{buras:06b,takiwaki:12}
and where $\Phi$ is the local gravitational potential and thus $ d
\Phi/ d r$ is the local gravitational acceleration. For simplicity, we
approximate the gravitational acceleration as $-GM(r) r^{-2}$ assuming
an angle-averaged spherical matter distribution in our postprocessing
analysis.   

\cite{foglizzo:06} pointed out that Eq.~(\ref{eq:ledoux}) is an
insufficient criterion for the development of large-scale convective
instability in the postshock region. A small (linear) perturbation
that could seed convection in the unstable gain layer is advected in
towards the convectively stable cooling layer with the background
flow. This advection may occur faster than the time it takes for
convection to grow from the small perturbation. It is thus necessary
to compare the advection timescale $\tau_\mathrm{adv}$ with the growth
time for convection in the gain layer, $\tau_\mathrm{conv} \approx
\omega_{BV}^{-1}$.
\cite{foglizzo:06} defined the quantity
\begin{equation}
\chi = \int_{R_\mathrm{gain}}^{R_\mathrm{shock}}
\frac{\omega_\mathrm{BV}}{|v_r|}
dr = \frac{\tau_\mathrm{adv}}{\tau_\mathrm{conv}} \,\,,
\label{eq:chi}
\end{equation}
where $v_r$ is the radial velocity through the gain region. A small
scale perturbation of magnitude $\delta_\mathrm{in}$ entering the gain
layer from above may at most grow by a factor $\exp{(\chi)}$ to
$\delta_\mathrm{out} = \delta_\mathrm{in} \exp{(\chi)}$ during its
advection through the gain layer \citep{scheck:08}. According to the
linear analysis of \cite{foglizzo:06}, $\chi \gtrsim 3$ is required
for convection to develop in the gain layer from small perturbations
$\delta_\mathrm{in}$. \cite{scheck:08} noted, then demonstrated, that
the situation is different if the seed perturbations
$\delta_\mathrm{in}$ are sufficiently large so that the time integral
of the buoyant acceleration becomes comparable to the advection
velocity. In this case, the advected seed may grow into a buoyant
plume and stay in the gain layer instead of leaving it. The results of
\cite{scheck:08} indicate that local seed perturbations of order
$1\%$, e.g., in the upstream radial velocity, may already be
sufficient to trigger convection even if $\chi < 3$.

\begin{figure}[t]
\centering
\includegraphics[width=0.95\columnwidth]{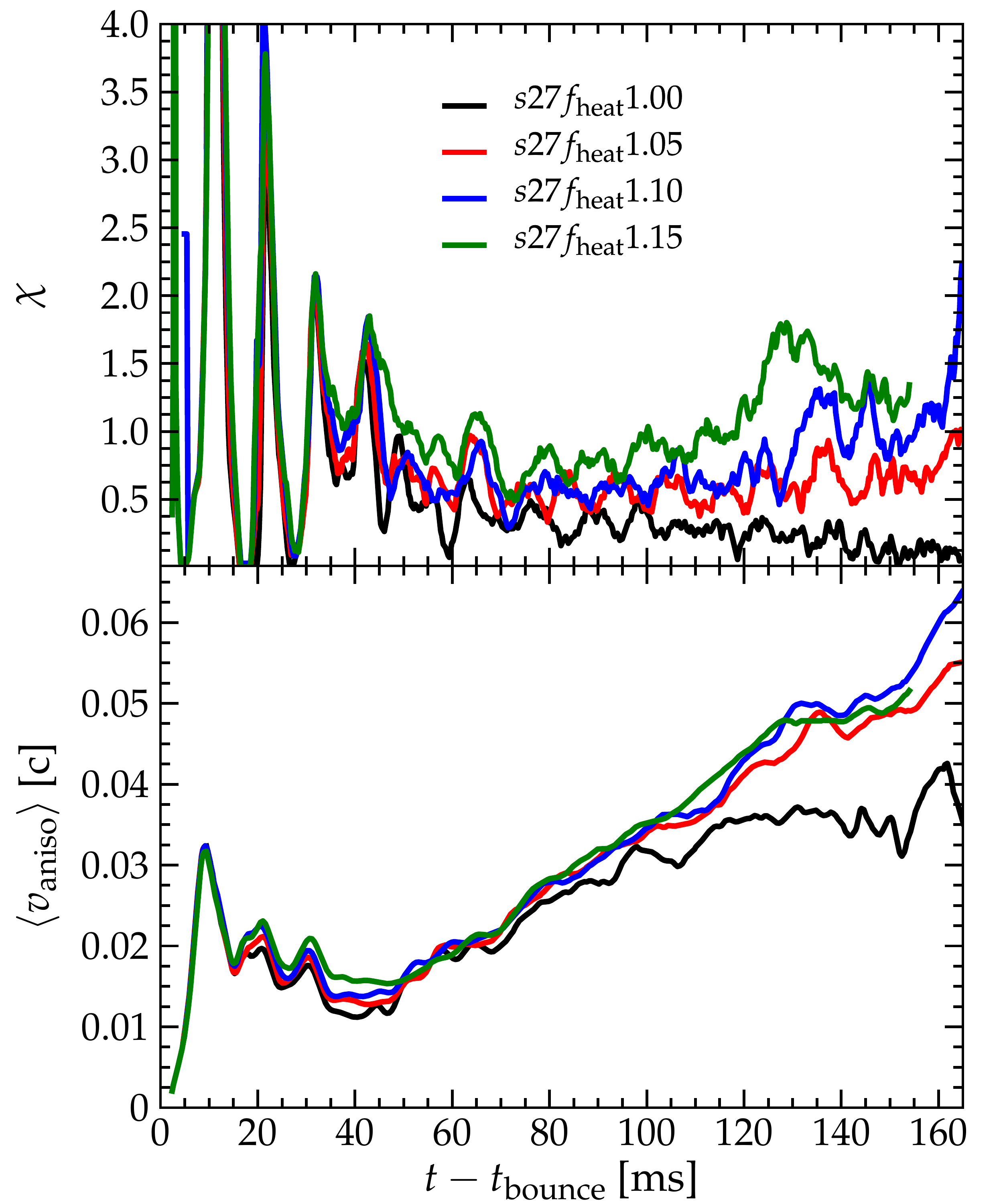}
\caption{{\bf Top panel:} Foglizzo parameter $\chi$ (Eq.~\ref{eq:chi})
  as a function of time after bounce. At times $\lesssim
  40\,\mathrm{ms}$ after bounce, the shock is still expanding and a
  quasi-stationary gain layer has not yet developed. $\chi$ is not
  reliable in that phase. At later times, it stays consistently below
  the critical value of $3$ suggested by \cite{foglizzo:06} as being
  necessary for convection to develop from arbitrarily small
  perturbations.  Note that stronger neutrino heating leads to greater
  $\chi$, since $\omega_\mathrm{BV}$ is larger.  {\bf Bottom panel:}
  Density-weighted average of the anisotropic velocity $\langle
  v_\mathrm{aniso} \rangle$ (Eq.~\ref{eq:vaniso}) in the gain layer
  inside the minimum shock radius. As in the case of $\chi$, this
  quantity is not reliable in the highly dynamic early postbounce
  phase. The early peak around $10\,\mathrm{ms}$ is related to prompt
  convection, which ebbs over $\sim$$30\,\mathrm{ms}$. Starting at
  $\sim$$40\,\mathrm{ms}$ after bounce, when neutrino driving becomes
  efficient (cf.~Fig.~\ref{fig:heat}), $\langle v_\mathrm{aniso}
  \rangle$ increases nearly monotonically in a very similar way in all
  models. Only model $s27f_\mathrm{heat}1.00$, which has the weakest
  neutrino heating, deviates from this trend at times
  $\gtrsim$$100\,\mathrm{ms}$ after bounce.}
\label{fig:chi_vaniso}
\end{figure}

If convection does develop, a simple measure of its strength is
the anisotropic velocity $v_\mathrm{aniso}$, which we define, following
 \cite{takiwaki:12}, as
\begin{equation}
\label{eq:vaniso}
v_\mathrm{aniso} = \sqrt{\frac{\left\langle \rho \left[ \left(v_r -\langle
    v_r \rangle_{4\pi} \right)^2 + v_\theta^2 + v_\varphi^2 \right]
  \right\rangle_{4\pi}}{ \langle \rho \rangle_{4\pi} }}\,\,,
\end{equation}
where $\langle . \rangle_{4\pi}$ denotes an angle average at fixed
radius. $v_\mathrm{aniso}$ essentially extracts the
  magnitude of the velocity component that does not belong to a purely
  radial background flow. We compute $v_\mathrm{aniso}$ by
introducing a spherical auxillary grid onto which we interpolate the
Cartesian coordinate velocity components and transform to obtain
$v_r$, $v_\mathrm{\theta}$, and $v_\mathrm{\varphi}$. We then
integrate over $4\pi$ steradian at each radius $r$ to obtain the
various angle-averaged quantities. $v_\mathrm{aniso}$ is high in
regions of large fluctuations in $v_r$ and high non-radial velocities
$v_\theta$ and $v_\varphi$. We note that high $v_\mathrm{aniso}$ in
the postshock region is a good measure for non-radial flow in that
region. If this non-radial flow is due to prompt/neutrino-driven
convection or induced by the SASI is difficult to decide, in
particular when the SASI has reached the non-linear
regime. $v_\mathrm{aniso}$ is thus most useful at early postbounce
times and both its time evolution and radial distribution must be
carefully considered.

In Fig.~\ref{fig:bv_vaniso}, we present colormaps showing the time
evolutions of radial profiles of the angle-averaged $\omega_{BV}$
(left panels) and $v_\mathrm{aniso}$ (right panels) for models
$s27f_\mathrm{heat}1.05$ (top panels) and $s27f_\mathrm{heat}1.15$
(bottom panels) as representative cases for moderate and strong
neutrino heating. The qualitative evolution is the same in all models.

Within milliseconds of bounce, a highly convectively unstable region
develops where the negative entropy gradient left behind by the
stalling shock is strongest. As is evident from the $v_\mathrm{aniso}$
colormaps, a strong burst of prompt convection develops and smoothes
out this entropy gradient within $\sim$$20\,\mathrm{ms}$ of
bounce. The highly dynamical early phase of shock expansion and prompt
convection is over by $\sim$$40\,\mathrm{ms}$ after bounce, when the
shock has settled at $\sim$$100-120\,\mathrm{km}$. At this time, the
gain layer has developed and neutrino heating creates a negative
entropy gradient and thus instability to convection between
$\sim$$80\,\mathrm{km}$ and the shock. Also, deleptonization at the
edge of the protoneutron star (cf.~Fig.~\ref{fig:profiles}) creates a
negative lepton gradient, driving protoneutron star convection, which
sets in at $35-40\,\mathrm{ms}$ after bounce and is clearly marked by
a band of high $v_\mathrm{aniso}$, spatially coinciding with the band
of convective instability in the protoneutron star. 

Figure~\ref{fig:chi_vaniso} shows the evolution of the Foglizzo
parameter $\chi$ (Eq.~\ref{eq:chi}; top panel) and the
density-weighted average anisotropic velocity in the gain layer
(bottom panel). At the early postbounce times at which prompt
convection takes place, both quantities are poorly defined, since
the shock expansion is still rather dynamic and a quasi-stationary
gain layer does not yet exist. This explains the large variations seen
at early times in particular in $\chi$. Once the postbounce
quasi-equilibrium in the postshock region is established, $\chi$
settles at values between $0$ and $2$ in all models, which is consistent
with what \cite{mueller:12b} found for the $s27$ progenitor.  

The $\chi \gtrsim 3$ criterion proposed by \cite{foglizzo:06} for the
development of convection in the gain layer is never fulfilled in any
of our models.  Nevertheless, neutrino-driven convection does develop
and becomes strong in all of our models. This is obvious from the
radial $v_\mathrm{aniso}$ distribution shown in
Fig.~\ref{fig:bv_vaniso}.  As soon as the gain layer develops and
$\omega_\mathrm{BV}$ becomes large, a broad region of high
$v_\mathrm{aniso}$ appears and traces the region of instability.  This
is indeed neutrino-driven convection, as can be seen from the entropy
slices in the top panel of Fig.~\ref{fig:visit_entropy}, which show
fully developed neutrino-driven convection at $\sim$$80\,\mathrm{ms}$
after bounce. The development of neutrino-driven convection in the
gain layer can also be inferred from the density-weighted average
$v_\mathrm{aniso}$ over the gain layer (bottom panel of
Fig.~\ref{fig:chi_vaniso}). $\langle v_\mathrm{aniso}\rangle$ has an
initial local maximum due to prompt convection and decreases as the
latter ebbs only to increase again at $\gtrsim40\,\mathrm{ms}$ after
bounce when neutrino heating in the gain layer becomes efficient and
drives convection (cf.~Fig.~\ref{fig:heat}).

\cite{mueller:12b}, in their axisymmetric simulation of the $s27$
progenitor, did not observe the development of neutrino-driven
convection, in agreement with the prediction of \cite{foglizzo:06}
that for $\chi \lesssim 3$ small perturbations are advected out of the
gain layer before they can grow into buoyant plumes. While there are
many technical differences between the simulation of
\cite{mueller:12b} and the ones presented here, the key difference
relevant for the development of neutrino-driven convection in our
simulations is our choice of a Cartesian AMR grid as opposed to the
spherical polar grid of the axisymmetric code of \cite{mueller:12b}.
A spherical polar grid is ideal for tracking the spherically-symmetric
collapse phase and the upstream flow outside the shock after bounce.
Seed perturbations remain minimal and neither prompt nor
neutrino-driven convection grow in the simulation of
\cite{mueller:12b}. Our Cartesian AMR grid, on the other hand, leads
to significant perturbations in multiple ways: (\emph{i}) The
Cartesian grid itself only imperfectly resolves spherical flow and
perturbations of at most order $dx / R_\mathrm{shock}$, where $dx$ is
one linear computational cell size, are generated locally at the shock
front. (\emph{ii}) Also due to its rectangular nature, the grid has
$\ell = 4, m = 4$ symmetry, which leads to buildup of numerical noise
primarily in modes with $\ell = 4, m = \{-4,0,4\}$. (\emph{iii}) In
our AMR setup, the shock is formed on the finest grid and then is
allowed to pass through two mesh refinement boundaries before it
reaches the grid that will track its subsequent evolution. The
crossing of AMR boundaries causes large perturbations in the shock
front that are also of $\ell = 4$ character. (\emph{iv}) The AMR grid
that tracks the shock front expands whenever the shock expands. The
AMR boundary must constantly be filled via interpolation from the next
coarser grid, which also introduces noise.  Points
(\emph{i})-(\emph{ii}) are true for any code using a Cartesian grid,
e.g., the \code{CASTRO} code used in the recent simulations of
\cite{burrows:12,murphy:12,dolence:12}, while (\emph{iii})-(\emph{iv})
are due to our particular approach in \code{Zelmani}, which may or may
not be different from what is done in \code{CASTRO} and other codes.

There are multiple ways in which one could quantify the magnitude of
the perturbations present in the early postbounce phase in our models.
One indicator may be the relative deviation of the shock front from
spherical symmetry quantified in Fig.~\ref{fig:modes}. The $\ell = 4$
grid modes indeed imprint themselves on the shock front, though the
deviation of the shock itself from sphericity is not large.  The
root-square-sum $A_4$ of the normalized $\ell = 4, m =
\{-4,\ldots,4\}$ components of the shock front has a maximum of
$\sim1.4\%$ at $\sim$$10\,\mathrm{ms}$ after bounce. This could be
interpreted as a lower bound on the deviation from sphericity of the
postshock flow and may already be sufficient to seed convection
\citep{scheck:08}.  Alternatively, we consider the relative
root-mean-square deviation from sphericity of any fluid quantity $X$
on a spherical shell of radius $R$,
\begin{equation}
\xi(X) = \frac{\sqrt{ \langle (X - \langle X\rangle_{4\pi})^2 \rangle_{4\pi} }}
{\langle X \rangle_{4\pi}}\,\,,
\end{equation}
where $\langle \cdot \rangle_{4\pi}$ denotes an angular average at
fixed radius and we have dropped the dependence on $R$ for simplicity.
Evaluating $\xi$ for density, radial velocity, entropy, and pressure
in the preshock region ($R > R_\mathrm{shock,max}$), we find only very
small deviations from sphericity of order $0.1\%$ at any time.  We
carry out this analysis also at a radius just inside the shock
(dynamically adjusting $R$ to be $\sim R_\mathrm{shock,min} -
1\,\mathrm{km}$), which should be reliable in the dynamical
shock expansion phase. Any perturbations would have to come from
shock passage, since convection had no time to grow. For this, we find
large deviations of $5-10\%$ in density, entropy, and
pressure\footnote{The deviation of the radial velocity is of order
  unity there, which is readily explained by the extreme variation of
  $v_{r}$ across the shock and is thus not a reliable
  measure.}. These deviations are present already milliseconds after
bounce and they peak when the shock passes through the boundary of the
second finest refinement level (at $59\,\mathrm{km}$) at
$\sim$$3\,\mathrm{ms}$ after bounce. This indicates that shock passage
through refinement boundaries may be the dominant source of numerical
perturbations in our simulations.

The large-amplitude perturbations present in the early postbounce flow
are more than sufficient to overcome advection and seed prompt
convection, which grows within milliseconds of bounce in our models
(cf.~Figs.~\ref{fig:bv_vaniso} and \ref{fig:chi_vaniso}).
Neutrino-driven convection is, in turn, seeded by the turbulent flow
of prompt convection and by additional, though much smaller magnitude,
noise coming from interpolation at the AMR boundary and from the
Cartesian representation of the spherically accreting
  outer core.

\subsubsection{SASI}

\begin{figure}[t]
\centering
\includegraphics[width=0.95\columnwidth]{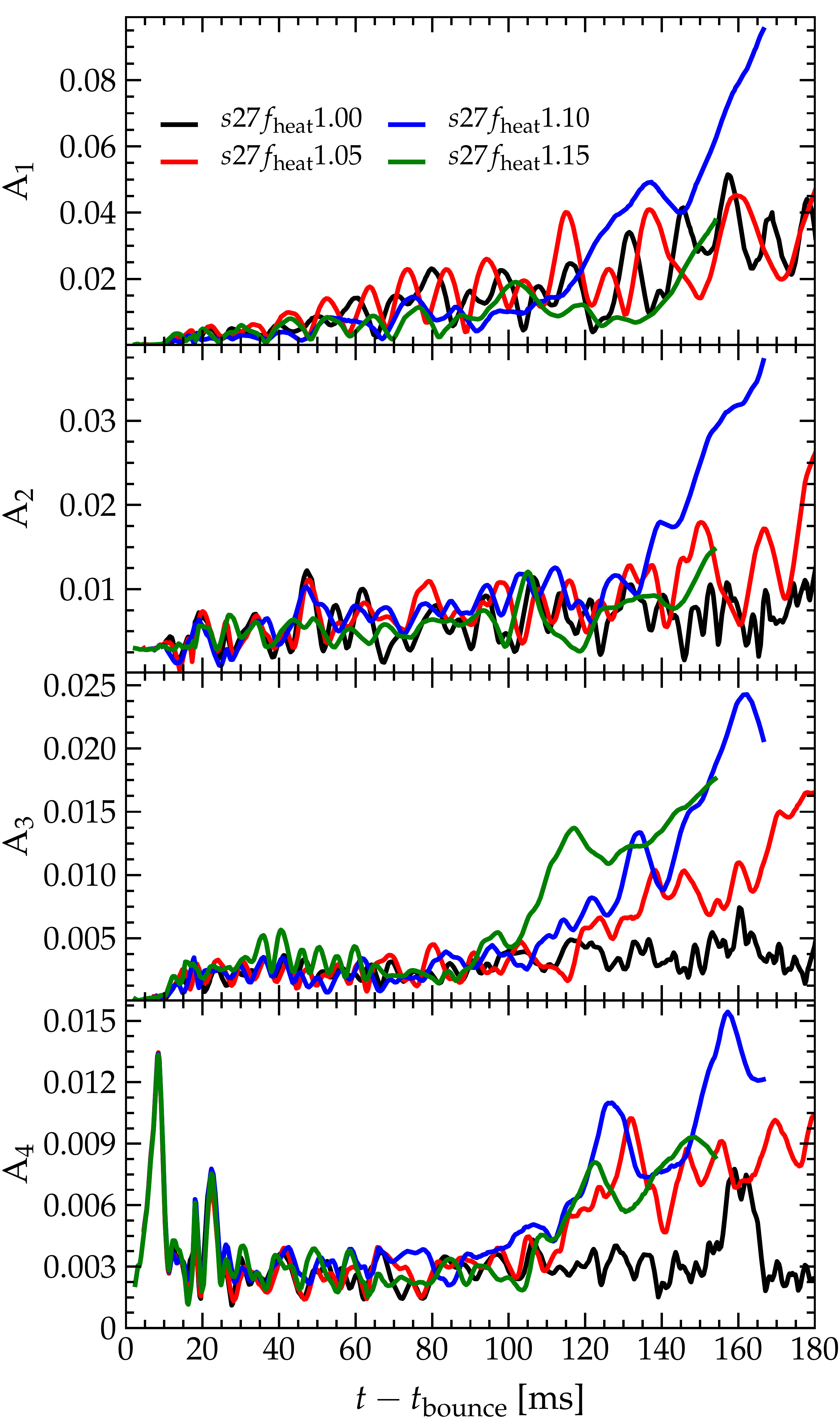}

\caption{Time evolution of the normalized root-square-summed spherical
  harmonic mode amplitudes of the shock in all models for each $\ell$
  in $\{1,2,3,4\}$: $A_1$, $A_2$, $A_3$, and $A_4$
  (Eq.~\ref{eq:Al}). Note the dominance of the $\ell = 4$
  perturbations from the Cartesian grid at early times.  The $A_1$
  amplitude becomes dominant $\sim 40-50\,\mathrm{ms}$ after bounce
  and shows oscillatory features in models $s27f_\mathrm{heat}1.00$
  and $s27f_\mathrm{heat}1.05$. Models $s27f_\mathrm{heat}1.10$ and
  $s27f_\mathrm{heat}1.15$ have no obvious oscillatory behavior of
  $A_1$, but develop large non-oscillatory amplitudes at late times
  when the shock in these models reaches large radii.}
\label{fig:modes}
\end{figure}

Convective overturn, first prompt, then neutrino-driven, develops
early on in our simulations and appears dominant. We can, however, not
yet exclude growth of the SASI\@. The conditions for SASI growth are
very different from those for convection. Any standing accretion shock
is unstable to the SASI, with $\ell = 1, m=0,\pm 1$ modes being the
most unstable and growing from arbitrarily small
  perturbations (e.g., \citealt{guilet:12}). The linear growth rate
of the SASI can be expressed as
\begin{equation}
\omega_\mathrm{SASI} = \frac{\ln |\mathcal{Q}|}{\tau_\mathrm{cyc}}\,,
\end{equation}
where $\mathcal{Q}$ is the cycle efficiency, defined as the
amplification factor of perturbations in each advective-acoustic
cycle, and $\tau_\mathrm{cyc}$ is the duration of a cycle (see, e.g.,
\citealt{scheck:08} for a detailed discussion). Qualitatively,
$\tau_\mathrm{cyc}$ depends on the radius at which the shock stalls
and on the timescale for advection of entropy/vorticity perturbations
between shock and protoneutron star edge. A smaller shock radius and
shorter advection time will thus lead to a smaller $\tau_\mathrm{cyc}$
and faster SASI growth. Strong neutrino heating, as pointed out by
\cite{yamasaki:07} and \cite{scheck:08}, increases the buoyancy in the
gain layer and leads to both larger $\mathcal{Q}$ and shock
oscillation frequencies (connected with $\tau_\mathrm{cyc}$), while
the growth rate is not strongly affected.

\begin{figure*}[t]
\centering
\includegraphics[width=0.95\columnwidth]{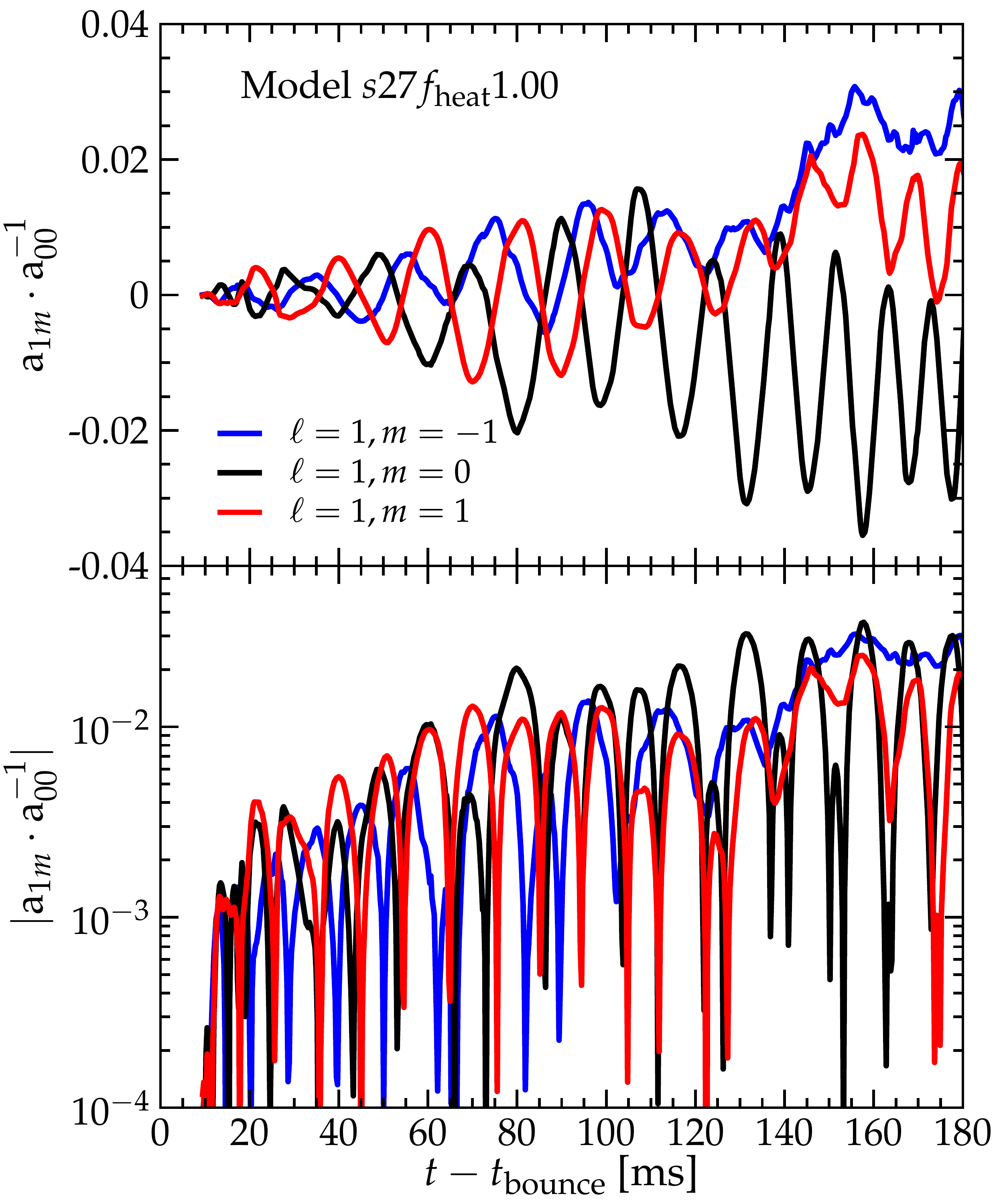}
\includegraphics[width=0.95\columnwidth]{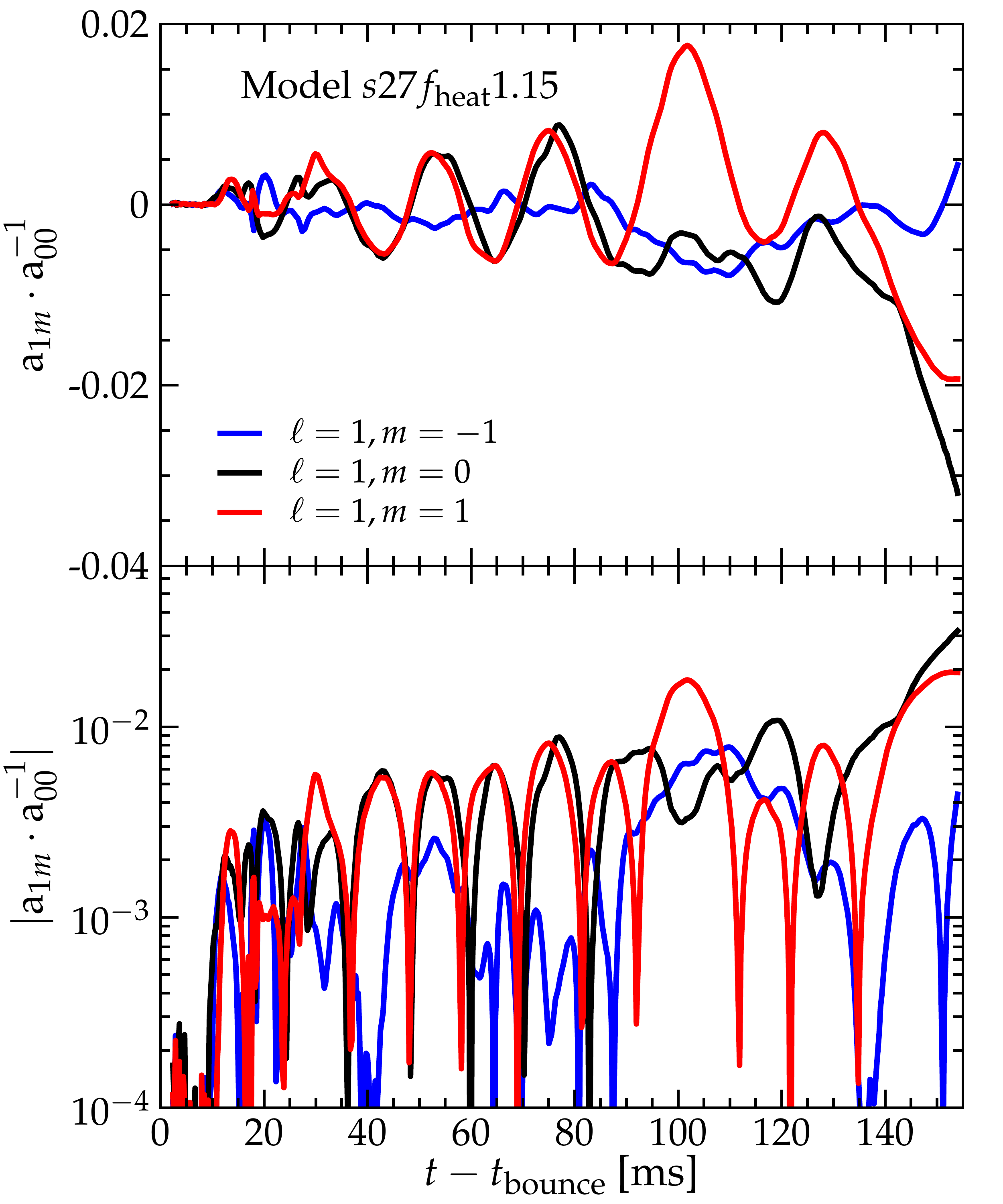}
\caption{Normalized $\ell = 1, m = \{-1,0,1\}$ mode amplitudes $a_{1
    m}/a_{00}$ of the shock front plotted on a linear scale (top
  panels) and their absolute values plotted on a logarithmic scale
  (bottom panels) for models $s27f_\mathrm{heat}1.00$ (left panels)
  and $s27f_\mathrm{heat}1.15$ (right panels). Model
  $s27f_\mathrm{heat}1.00$ shows a clear exponential growth of
  oscillatory modes, but saturation occurs at amplitudes that are
  about an order of magnitude smaller than in the 2D simulation of the
  same progenitor carried out by \cite{mueller:12b}. Model
  $s27f_\mathrm{heat}1.15$, which has strong neutrino heating and
  intense neutrino-driven convection also shows some oscillatory $\ell
  = 1$ mode growth, though at a longer oscillation period, lower
  saturation amplitudes, and without a well defined exponential growth
  phase.}
\label{fig:a1s}
\end{figure*}

A characteristic feature of the SASI in its linear phase
  is the exponential growth of oscillatory low-mode deformations of
  the shock front. We look for evidence for the SASI in our
simulations by decomposing the shock surface
$R_\mathrm{shock}(\theta,\phi)$ into spherical harmonics: 
\begin{equation}
\label{eq:alm}
  a_{\ell m} = \frac{(-1)^{|m|}}{\sqrt{4\pi(2\ell+1)}} \int_{4\pi} R_\mathrm{shock}
  (\theta,\phi) Y^m_\ell (\theta,\phi) d\Omega \ .
\end{equation}
Note that $a_{00}$ corresponds to the average shock radius and that
the definition of the $a_{\ell m}$ used here gives individual $a_{\ell
  m}$ amplitudes that are a factor of $(2\ell + 1)$ smaller than the
definition for $a_{\ell 0}$ used by \cite{mueller:12b} in the
axisymmetric case, but at each $\ell$, there are $(2\ell+1)$ more
modes in our case.  The $Y^m_\ell$ are the standard real spherical
harmonics (e.g., \citealt{boas:06}), which we use with the
normalization factors given in \cite{burrows:12}.  We also define the
quantities $A_\ell$ as the root-square-sum of the $a_{\ell m}$ for a
given $\ell$ normalized by the average shock radius $a_{00}$,
\begin{equation}
A_\ell = \frac{1}{a_{00}} \sqrt{ \sum_{m=-\ell}^\ell a^2_{\ell m} } \,\,.
\label{eq:Al}
\end{equation}

In Fig.~\ref{fig:modes}, we present in four panels, from top to
bottom, the time evolutions of the $A_1 - A_4$ amplitudes of the shock
front in all four models. In the first $\sim$$20\,\mathrm{ms}$ after
bounce, the initial $\ell = 4$ deformation due to our Cartesian grid
imprints itself onto the shock front and the $A_4$ amplitude is
dominant. Subsequently, the other modes grow. For SASI growth, the
expectation is that the $\ell =1, m=\{-1,0,1\}$ modes have the fastest
growth rate and have oscillatory behavior, which should be reflected
in the $A_1$ amplitude. In models $s27f_\mathrm{heat}1.00$ and
$s27f_\mathrm{heat}1.05$, $A_1$ indeed is the fastest growing
amplitude and shows the expected oscillatory behavior throughout the
simulated postbounce interval, suggesting the presence of the SASI\@.
However, the maximum value of $A_1$ reached is $\sim$$0.04$, which is
an order of magnitude smaller than what was reported by
\cite{mueller:12b} for their 2D simulation of the $s27$
progenitor. 

For the two models with stronger neutrino heating,
$s27f_\mathrm{heat}1.10$ and $s27f_\mathrm{heat}1.15$, the situation
is different. Their $A_1$ and $A_2$ amplitudes hover around very
similar small values without obvious oscillatory behavior until
$\sim$$100\,\mathrm{ms}$ after bounce, when large-scale deviations
from sphericity (cf.~Fig.~\ref{fig:volume}) lead to strongly growing
amplitudes in all $\ell$.  This was also observed in the
high-luminosity light-bulb simulations of \cite{burrows:12} and
\cite{dolence:12}. $A_1$ is the dominant amplitude and reaches
$\sim$$0.1$ in model $s27f_\mathrm{heat}1.10$ and about $0.03$ in
model $s27f_\mathrm{heat}1.15$ at the end of its simulation. It is
interesting to note that model $s27f_\mathrm{heat}1.05$, which has a
positively trending shock radius at the end of its simulation, has
clearly growing $A_2$, $A_3$, and $A_4$ amplitudes at late times,
while $A_1$ remains the dominant mode with stable amplitudes near
$0.03$.

For further insight into the nature of the observed mode evolution, we
plot, in Fig.~\ref{fig:a1s}, the individual $\ell = 1, m=\{-1,0,1\}$
normalized mode amplitudes $a_{1m} / a_{00}$ in linear (top panels)
and logarithmic scale (bottom panels) for models
$s27f_\mathrm{heat}1.00$ (left panels) and $s27f_\mathrm{heat}1.15$
(right panels).  The former model has the weakest neutrino heating and
least vigorous neutrino-driven convection of all our models while the
latter model has the strongest heating and most vigorous
convection. All $\ell = 1$ modes in model $s27f_\mathrm{heat}1.00$
show a clear oscillatory behavior and, importantly, an exponential
growth phase between $\sim$$20$ and $\sim$$80\,\mathrm{ms}$ after
bounce can be made out. However, saturation occurs at low $a_{1m} /
a_{00} \sim 0.01$ for all modes. As noted before, this is an order of
magnitude smaller than found in the axisymmetric simulations of
\cite{mueller:12b} (in which neutrino-driven convection did not
develop as a primary instability).

Interestingly, some of the $a_{1m} / a_{00}$ modes in model
$s27f_\mathrm{heat}1.15$ do exhibit oscillatory behavior, though with
larger periods than in model $s27f_\mathrm{heat}1.00$. This is
expected for SASI growth under the influence of strong neutrino
heating \citep{yamasaki:07,scheck:08}. The growth also saturates more
quickly at amplitudes that remain a factor of $\sim$$2$ smaller than
in model $s27f_\mathrm{heat}1.00$ until $\sim$$100\,\mathrm{ms}$ after
bounce, when the mode growth becomes non-oscillatory. It is not
possible to unambiguously and clearly identify a phase of exponential
growth of the $a_{1m} / a_{00}$ modes in this model.

In summary, there is clear evidence for SASI growth in our models. It
is strongest in the model with the least neutrino heating and weakest
neutrino-driven convection. It is weakest in the model with the most
neutrino heating and the strongest neutrino-driven
convection. However, even in the model in which SASI growth is
strongest, the SASI saturates at amplitudes that are an order of
magnitude smaller than in the 2D simulation of \cite{mueller:12b},
which did not have any neutrino-driven convection. These observations
suggest that 3D neutrino-driven convection is indeed detrimental to
the development of large-amplitude SASI\@. This confirms the findings
of \cite{scheck:08,guilet:10,burrows:12,dolence:12}. Furthermore, our
results show that both instabilities can coexist and grow at the same
time, but even if convection is suppressed (a case we cannot study in
our 3D Cartesian AMR code), the nearly equal splitting of the $\ell =
1$ power across the three azimuthal $m$ modes in 3D, will likely
reduce the magnitude of deviations from sphericity that can be driven
by the SASI alone. Moreover, the SASI, once it has reached its
non-linear phase, will trigger neutrino-driven convection
\citep{scheck:08,guilet:10,burrows:12,mueller:12b}, which may very
well become the dominant instability, in particular if neutrino
heating is strong.

\subsection{Criteria for Neutrino-Driven Explosions}
\label{sec:criteria}

The simulations presented here end before an explosion is fully
developed in any of our models. Nevertheless, interesting trends can
be observed. Models $s27f_\mathrm{heat}1.10$ and
$s27f_\mathrm{heat}1.15$ have strongly positively trending shock radii
at the end of their simulations. The shock in model
$s27f_\mathrm{heat}1.05$ also expands at late times, but the
development of an explosion is definitely more marginal. Model
$s27f_\mathrm{heat}1.00$ has a receding shock and thus a rather
negative prognosis regarding explosion.

A variety of criteria for neutrino-driven explosions have been
discussed in the literature and it is interesting to see how the
trends observed in our models compare with what is expected from
theory and other simulation results. 

From the bottom panel of Fig.~\ref{fig:heat} we find that models with
stronger neutrino heating and, thus, more vigorous neutrino-driven
convection have systematically more mass in the gain layer
($M_\mathrm{gain}$) that can absorb neutrino energy. The low-amplitude
SASI seen in our models, which is strongest in models with weakest
heating and convection, does not appear to have any positive effect on
$M_\mathrm{gain}$ in our simulations. In models that are trending
towards explosion, $M_\mathrm{gain}$ increases as shock expansion sets
in. This is consistent with
previous work giving the most optimistic prognosis for models with the
greatest $M_\mathrm{gain}$ (e.g.,
\citealt{murphy:08,scheck:08,mueller:12a,hanke:12}).

Also shown in the bottom panel of Fig.~\ref{fig:heat} is the
density-weighted average of the specific entropy in the gain layer
($\langle s_\mathrm{gain}\rangle$). All models, trending towards
explosion or not, exhibit the same $\langle s_\mathrm{gain}\rangle$
evolution until $\sim$$130\,\mathrm{ms}$ after bounce, when the most
optimistic models actually move to somewhat smaller $\langle
s_\mathrm{gain}\rangle$ (cf.\ the discussion in \S\ref{sec:overall}).
Thus, in agreement with \cite{hanke:12}, the average
entropy in the gain layer is not a good indicator for a model's potential
for explosion.

\begin{figure}[t]
\centering
\includegraphics[width=0.95\columnwidth]{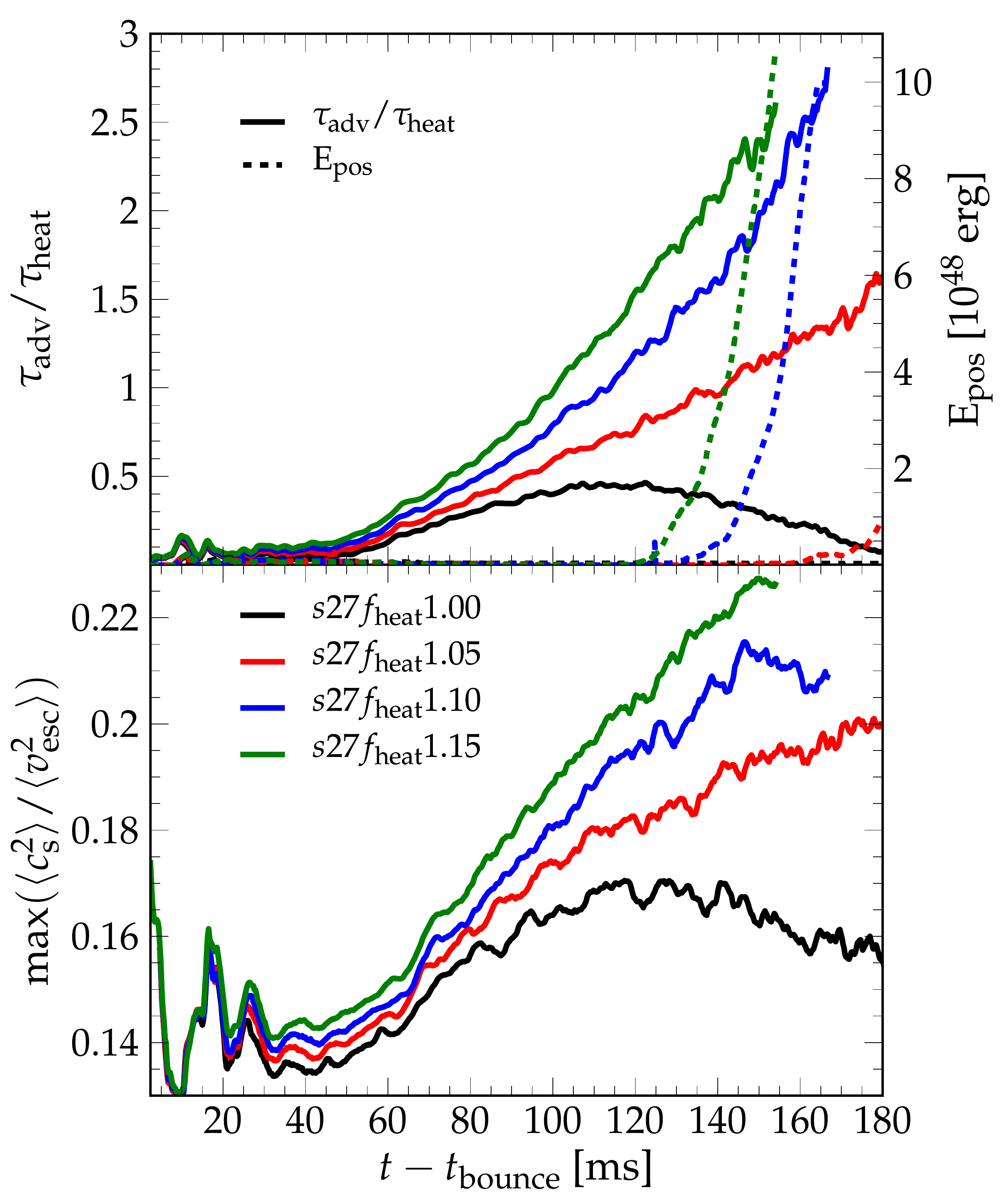}
\caption{{\bf Top panel:} Ratio of advection and heating timescales
  $\tau_\mathrm{adv}/\tau_\mathrm{heat}$ as a function of time after
  bounce in our models (left ordinate) and $E_\mathrm{pos}$, the
  volume integral over positive values of the total specific energy in
  the gain region (right
  ordinate). $\tau_\mathrm{adv}/\tau_\mathrm{heat} \gtrsim 1$ is
  considered to be a condition for runaway explosion. It is satisfied
  by all of our models with optimistic outlook. Models
  $s27f_\mathrm{heat}1.15$, $s27f_\mathrm{heat}1.10$, and
  $s27f_\mathrm{heat}1.15$ reach the threshold at $\sim$$100$,
  $\sim$$115$, and $\sim$$142\,\mathrm{ms}$ after bounce,
  respectively. Roughly $\sim$$20\,\mathrm{ms}$ later, these models
  are beginning to develop regions with positive total energy, which
  may be interpreted as the onset of explosion. {\bf Bottom panel:}
  Maximum of the ratio of the angle-averaged squared speed of sound to
  the angle-averaged squared escape velocity. According to the
  antesonic condition of \cite{pejcha:12a}, no solution for a
  spherical stationary accretion shock exists for $\max( c^2_s /
  v_\mathrm{esc}^2 ) > 3/16 \approx 0.19$ and an explosion is expected
  to set in in models that surpass this value.}
\label{fig:expl_criteria}
\end{figure}

A criterion frequently used to diagnose neutrino-driven explosions
arises from the comparison of the timescale for neutrino heating
$\tau_\mathrm{heat}$ and the advection timescale $\tau_\mathrm{adv}$
for material to pass throught the gain layer
\citep{burrows:93,janka:01,thompson:05,murphy:08}. If heating is
faster than advection through the gain layer, then a fluid parcel
entering the gain region may absorb sufficient energy to reach
positive total specific energy and thus become unbound. For
$\tau_\mathrm{adv} / \tau_\mathrm{heat} \gtrsim 1$, shock expansion
should set in, further increasing $\tau_\mathrm{adv}$ and thus leading
to positive feedback and runaway expansion.

In our simplified analysis, we set $\tau_\mathrm{heat} =
|E_\mathrm{gain}|/Q_\mathrm{net}$, where $Q_\mathrm{net}$ is the net
integral heating rate in the gain layer and $|E_\mathrm{gain}|$ is the
volume integral of the (Newtonian) total specific energy of material
in the gain layer, given, e.g., by the integral over
  Eq.~(3) of \citealt{mueller:12a}. We note that the internal energy
  of the LS220 EOS is defined with respect to a free neutron gas, this
  defines the zero of our internal energy. There are a variety of
possible definitions for $\tau_\mathrm{adv}$ (cf.~the discussions in
\citealt{murphy:08,marek:09,mueller:12a}). Here, we use the definition
$\tau_\mathrm{adv} = \dot M / M_\mathrm{gain}$, where
$M_\mathrm{gain}$ is the mass in the gain region and $\dot M$ is the
accretion rate through the shock. Note that this definition is
different from what we use in the computation of the Foglizzo $\chi$
parameter (Eq.~\ref{eq:chi}).

In the top panel of Fig.~\ref{fig:expl_criteria}, we plot
$\tau_\mathrm{adv} / \tau_\mathrm{heat}$ as a function of time after
bounce for all of our models (left ordinate).  The behavior is as
expected: the two models $s27f_\mathrm{heat}1.15$ and
$s27f_\mathrm{heat}1.10$, which are strongly trending towards
explosion reach $\tau_\mathrm{adv} / \tau_\mathrm{heat} \gtrsim 1$
already at $\sim$$100$ and $\sim$$115$ ms after bounce.  The marginal
model $s27f_\mathrm{heat}1.05$ also shows increasing
$\tau_\mathrm{adv} / \tau_\mathrm{heat}$, which reaches $1$ at $\sim
142$ ms after bounce. There is, however, no hope for model
$s27f_\mathrm{heat}1.00$, where $\tau_\mathrm{adv} /
\tau_\mathrm{heat}$ always remains below $\sim 0.5$.

Also shown in the top panel of Fig.~\ref{fig:expl_criteria} is
$E_\mathrm{pos}$ (right ordinate), the integral energy of unbound
material (with positive total specific energy, again defining the
internal energy with respect to a free neutron gas).  When
$\tau_\mathrm{adv}/\tau_\mathrm{heat} >1.4$ in our models, material
starts to become unbound and $E_\mathrm{pos}$ grows rapidly. However,
at the end of our simulations, it is still far away from the energy
needed to unbind the entire envelope and lead to a canonical $\sim$$1
B$ core-collapse supernova explosion.  We caution the reader to not
overinterpret $E_\mathrm{pos}$ -- it is unreliable at this
point. Rather, what is important to note is that towards the end of
the simulations there is an increasing amount of unbound material for
the highest values of $f_\mathrm{heat}$.  To obtain a quantitatively
reliable measure of the asymptotic explosion energy one must follow
the explosion to late times, consistently track or account for
recombination ($\sim$8-9\,MeV per nucleon), and consider the binding
energy of the overlying envelope ($\sim$\,1\,B; \citealt{whw:02}).

Finally, in the bottom panel of Fig.~\ref{fig:expl_criteria}, we plot
the time evolution of the maximum of the ratio of the angle-averaged
square of the speed of sound $\langle c_s^2 \rangle$ to the
angle-averaged square of the escape velocity, which we approximate as
$\langle v^2_\mathrm{esc} \rangle \approx 2 G M(r)/r$, where $M(r)$ is
the enclosed baryonic mass. This ratio is interesting, since
\cite{pejcha:12a} have recently derived the antesonic condition,  
\begin{equation} \max \left( \frac{c_s^2}{v_\mathrm{esc}^2} \right) >
  \frac{3}{16} \approx 0.19\,\,,
\end{equation}
beyond which no solution for a stationary spherically symmetric
accretion shock exists, marking the transition to explosion.  While
the expanding shocks in our models are far away from sphericity, we
find values of $\max(\langle c_\mathrm{s}^2 \rangle / \langle
v_\mathrm{esc}^2 \rangle) \gtrsim 0.2-0.22$, which is consistent with
the expectation of \cite{pejcha:12a}. Model $s27f_\mathrm{heat}1.00$,
which has the most pessimistic outlook, does not reach $\max(\langle
c_\mathrm{s}^2 \rangle / \langle v_\mathrm{esc}^2 \rangle) \gtrsim
0.19$, while the marginal model, $s27f_\mathrm{heat}1.05$ does. The
prognosis, according to \cite{pejcha:12a} is thus similar to the one
based on the $\tau_\mathrm{adv} / \tau_\mathrm{heat} > 1$ runaway condition.

\subsection{Gravitational Wave Signals}
\label{sec:gws}

\begin{figure*}[t]
\centering
\includegraphics[width=0.95\columnwidth]{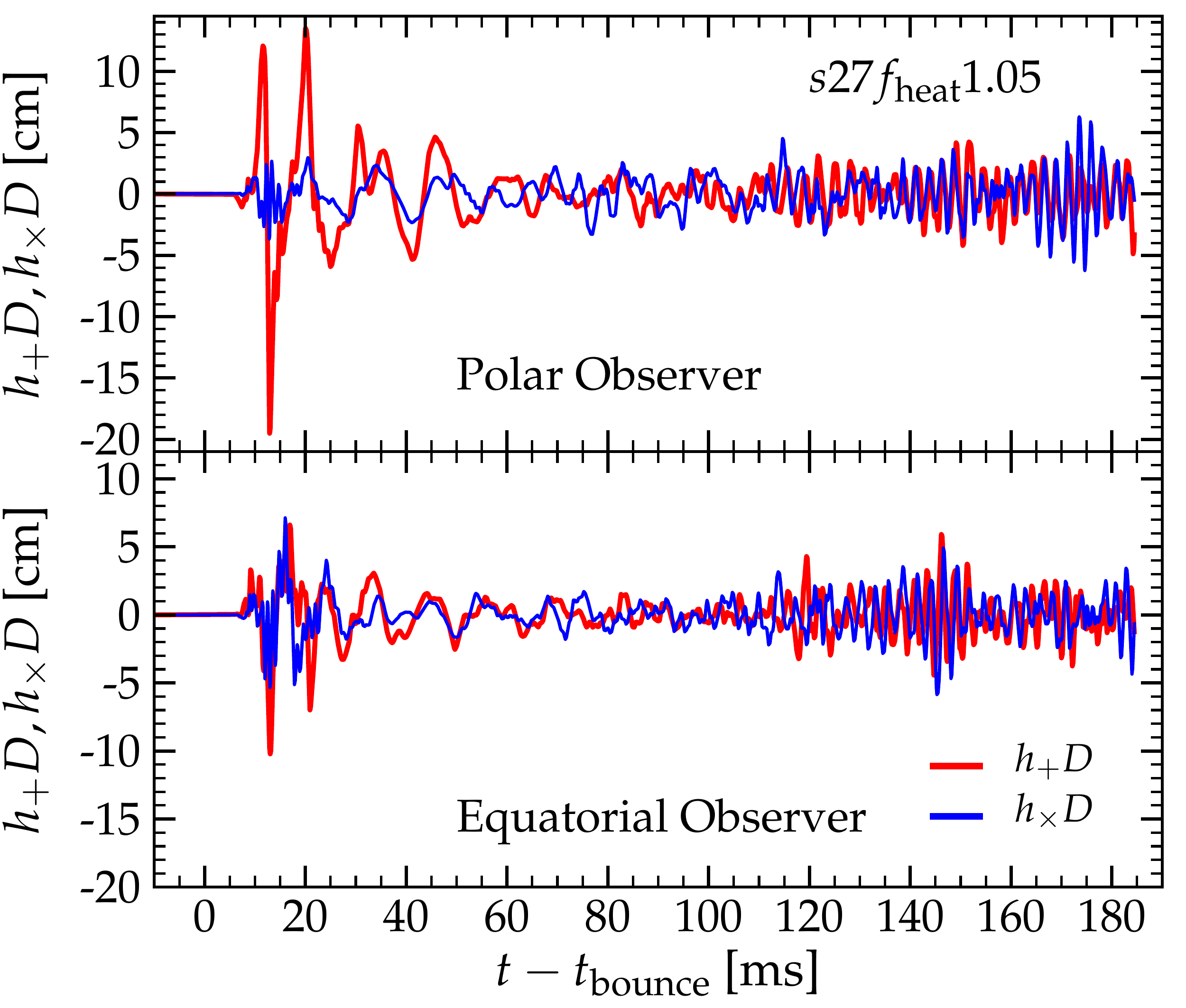}
\hspace*{0.5cm}
\includegraphics[width=0.95\columnwidth]{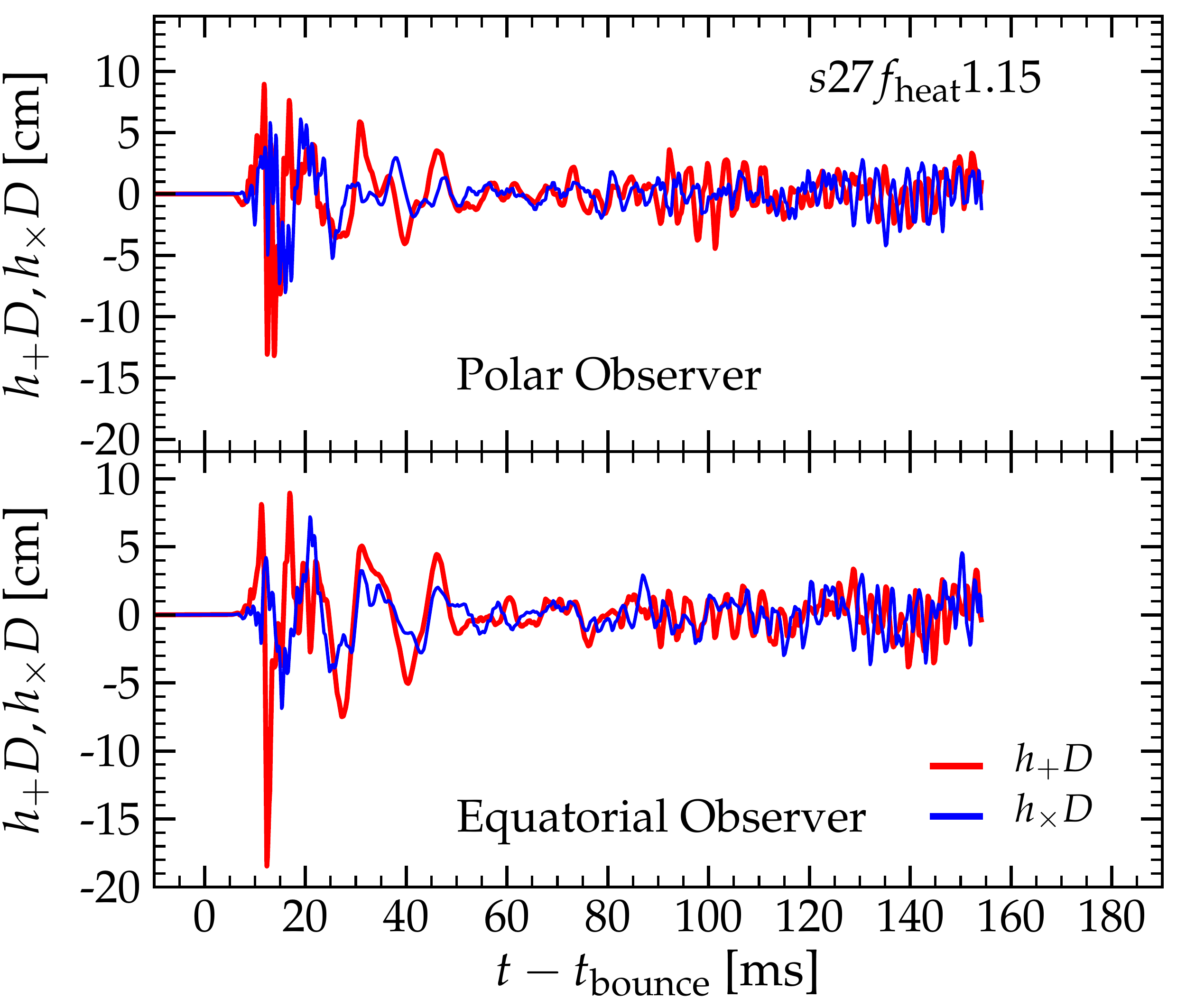}
\caption{{\bf Left panel}: Gravitational wave polarizations $h_+D$ and
  $h_\times D$ (rescaled by distance $D$) of model
  $s27f_\mathrm{heat}1.05$ as a function of postbounce time seen by
  and observer on the pole ($\theta = 0, \varphi = 0$; top panel) and
  on the equator ($\theta = \pi/2, \varphi=0$; bottom panel). {\bf
    Right panel}: The same for model $s27f_\mathrm{heat}1.15$. Both
  models show a burst of gravitational waves associated with
  large-scale prompt convection developing shortly after
  bounce. Subsequently, gravitational wave emission comes from
  aspherical flow in the gain layer, in the outer protoneutron star,
  and from descending plumes of material that are decelerated at the
  edge of the protoneutron star. The gravitational wave signals are
  trending towards higher frequencies with time.}
\label{fig:gw}
\end{figure*}

Besides the neutrino signals already discussed in
\S\ref{sec:checkleak}, gravitational waves (GWs) are the only other direct
probe of the processes occurring in the postshock region and in the
protoneutron star. The overall GW signature of
core-collapse supernovae has been reviewed in detail by \cite{ott:09}
and \cite{kotake:11b} and we refer the interested reader to these
reviews for an in-depth discussion of the various potential
GW emission processes and their underlying physics.

GW observations of the next galactic core-collapse
supernova could provide important insight into the role and relevance
of multi-dimensional fluid instabilities, rotation, the structure of
the protoneutron star, and the nuclear EOS
\citep{dimmelmeier:08,marek:09b,yakunin:10,murphy:09,roever:09,ott:09,kotake:11b}. Recently,
\cite{logue:12} carried out a proof-of-principle study, demonstrating
that Bayesian inference allows to select between different explosion
mechanisms for a galactic core-collapse supernova. The reliability of
this depends on the availability of robust waveform predictions from
simulations. Most currently available core-collapse supernova
waveforms come from 2D simulations (as summarized by
\citealt{ott:09,kotake:11}), which can predict only one of the two
independent polarizations. In the context of nonrotating or slowly
rotating neutrino-driven core-collapse supernovae, only very few
waveform predictions from 3D simulations without symmetry constraints
exist. \cite{fryer:04}, carried out Newtonian 3D smoothed-particle
hydrodynamics simulations with gray flux-limited diffusion neutrino
transport and studied the GW emission from matter
motions and asymmetric neutrino emission up to $\sim$$80\,\mathrm{ms}$
after bounce in a variety of different precollapse configurations with
and without initial rotation and large-scale
asphericities. \cite{kotake:09,kotake:11} performed Newtonian 3D
hydrodynamic simulations with a light-bulb scheme (similar to MB08, but
with a better approximation to changes in $Y_e$). They used analytic
initial conditions, a fixed accretion rate and a fixed inner spherical
boundary at $50\,\mathrm{km}$, but were able to evolve for
$\sim$$500\,\mathrm{ms}$ and studied the GW emission
from matter dynamics and asymmetric neutrino
emission. \cite{scheidegger:10} performed full 3D Cartesian (without
inner boundary) Newtonian collapse and postbounce simulations of a
slowly rotating progenitor with neutrino leakage (but no
heating). They employed a monopole approximation for gravity with
relativistic corrections and evolved to $\sim$$100\,\mathrm{ms}$ after
bounce. Recently, \cite{mueller:e12} presented Newtonian 3D postbounce
simulations with GR corrections to the monopole term of the Newtonian
potential. They used a time-dependent inner boundary that contracts
from $60-80\,\mathrm{km}$ to $15-25\,\mathrm{km}$ over $1\,\mathrm{s}$
following the prescription of \cite{scheck:08}, but were able to
evolve multiple progenitor models for $\gtrsim 1.2\,\mathrm{s}$ using
a ray-by-ray gray two-species approximate transport scheme (neglecting
$\nu_x$) and imposed neutrino luminosities at the inner boundary.
They extracted and studied in detail the GW emission due
to matter dynamics and anisotropic neutrino emission.

\begin{figure}
\centering
\includegraphics[width=0.95\columnwidth]{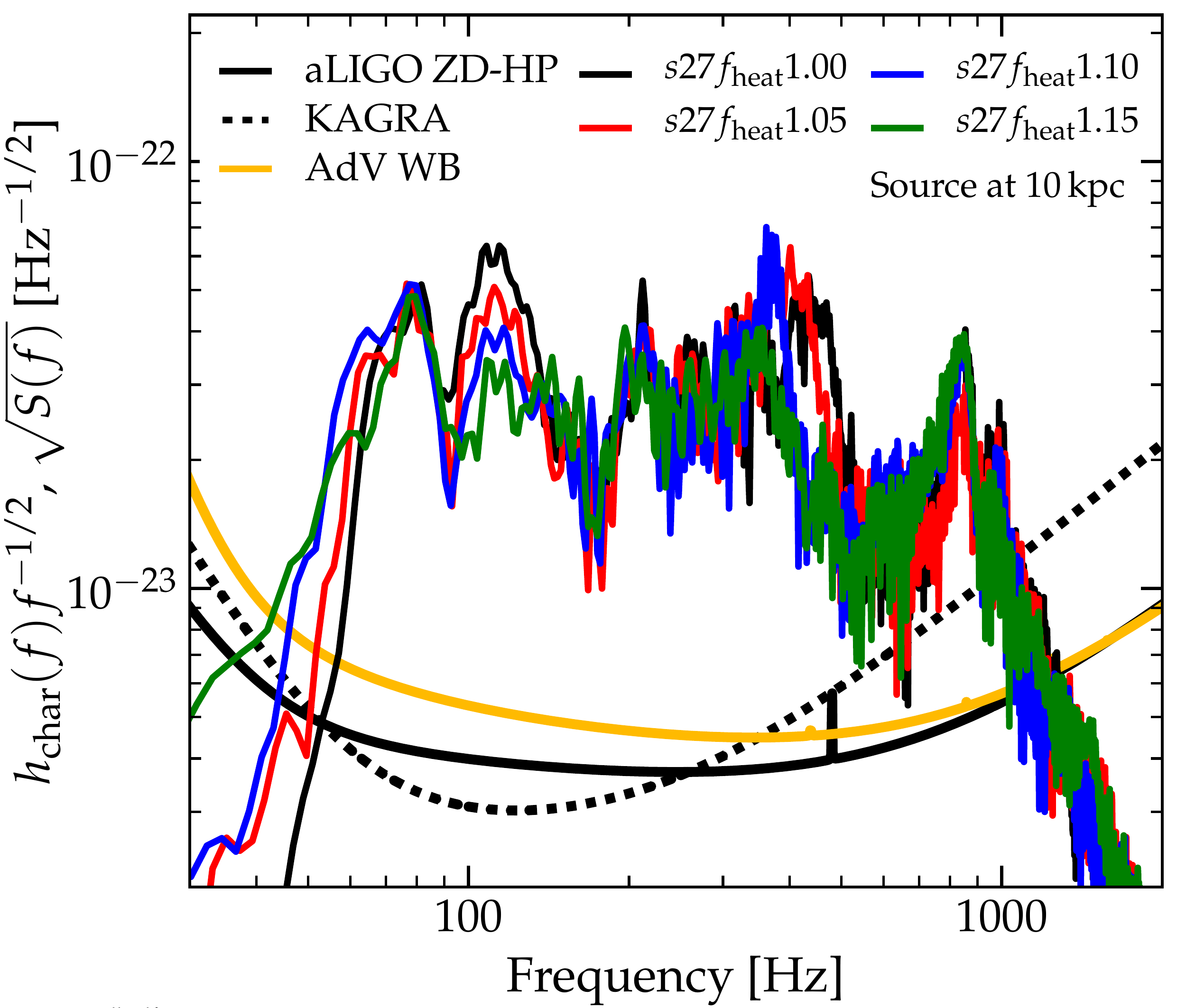}
\caption{Characteristic spectral strain spectra
  $h_\mathrm{char}(f) f^{-1/2}$ of all four models at a distance of
$10\,\mathrm{kpc}$ compared with the
  design noise levels $\sqrt{S(f)}$ of Advanced LIGO in the broadband
  zero-detuning high-power mode (aLIGO ZD-HP), KAGRA, and Advanced
  Virgo in wideband mode (AdV WB). }
\label{fig:gwspect}
\end{figure}

While the simulations presented in this study do not have the more
sophisticated neutrino transport treatment of \cite{mueller:e12}, they
do not have an artificial inner boundary with imposed core neutrino
luminosities, are carried out in full GR, and include the cooling due
to $\nu_x$ emission from the protoneutron star. It is, hence,
worthwhile to study the GWs emitted by our models. We restrict
ourselves to GWs from the dominant accelerated quadrupole matter
motions and ignore GWs from asymmetric neutrino emission. The
rationale for the latter is that our simple leakage scheme is unfit to
give a reasonable estimate for the true neutrino radiation field
anisotropy leading to GW emission. Moreover, as demonstrated by
previous work
\citep{kotake:09,kotake:11,mueller:e12,marek:09b,yakunin:10}, GW
emission due to asymmetric neutrino emission occurs at too low
frequencies to be relevant for earthbound detectors such as Advanced
LIGO \citep{harry:10,LIGO-sens-2010}, Advanced Virgo
\citep{virgostatus:11}, and KAGRA \citep{kagra:12}.

We employ the quadrupole approximation for extracting
GWs from our simulations and use the expressions
detailed in \cite{ott:12a}. In principle, we could extract the
gravitational waveforms directly from the spacetime, but the results
of \cite{reisswig:11ccwave} suggest that the quadrupole approximation
is very likely sufficiently accurate for stellar collapse spacetimes
with a protoneutron star. The full observer-angle independent 
GW signals for all models are available for download from \url{http://www.stellarcollapse.org/gwcatalog}\,.

In Fig.~\ref{fig:gw}, we plot the $h_+$ and $h_\times$ polarizations
of the GW signal (rescaled by distance $D$) for model
$s27f_\mathrm{heat}1.05$ (left panel) and model
$s27f_\mathrm{heat}1.15$ (right panel) as seen by observers on the
north pole ($\theta = 0, \varphi=0$; top panels) and on the equator
($\theta = \pi/2, \varphi = 0$; bottom panels). The GW signals emitted
by the other models are very similar and not shown. The early emission
sets in $\sim$$10\,\mathrm{ms}$ after bounce and is due to prompt
convection that dominates the aspherical dynamics in the early
postbounce phase, but has decayed by $\sim$$40\,\mathrm{ms}$ after
bounce. The GW signal from convection and other fluid instabilities is
of stochastic nature (cf.~\citealt{kotake:09,ott:09}) and its time
series cannot be predicted exactly. The GW signal of prompt convection,
since it is emitted within milliseconds of bounce by the strongest 
first few overturn cycles, is particular sensitive to the perturbations
seeding prompt convection. Note that the time series of $h_+$
and $h_\times$ from prompt convection in the two models are quite
different, but the overall amplitudes agree well, but peak in
different viewing directions.  The subsequent evolution of the GW
signals is similar in both models, both polarizations, and both
observer positions. After an intermittent quiescent phase, GW emission
picks up again at times $\gtrsim$$80\,\mathrm{ms}$ after bounce when
aspherical dynamics becomes strong throughout the entire postshock
region (cf.~Fig.~\ref{fig:bv_vaniso}). In this phase, the GW emission
transitions to higher frequencies, indicating that emission from
deceleration of downflows at the steep density gradient at the edge of
the protoneutron star (as first pointed out by \citealt{murphy:09})
and convection in the protoneutron star play an increasing role.
While both models have expanding shocks at the end of their
simulations, the shock acceleration has not become sufficiently strong
to lead to an offset in the GW signal (GW memory) seen in other
work that followed exploding models to later times (e.g.,
\citealt{murphy:09,yakunin:10,mueller:e12,kotake:09,kotake:11}).

The peak GW strain amplitudes reached in our models are from prompt
convection and go up to $|h|D\sim$$20\,\mathrm{cm}$
($\sim$$6.5\times10^{22}$ at
$10\,\mathrm{kpc}$). \cite{scheidegger:10} found
$|h|D\sim$$10\,\mathrm{cm}$ and \cite{fryer:04} found
$|h|D\sim$$12\,\mathrm{cm}$, but we note that the GW signal will
depend on the strength of prompt convection, which is different from
model to model. The approaches of \cite{mueller:e12} and
\cite{kotake:09,kotake:11} do not allow them to study prompt
convection. The typical amplitudes reached in the preexplosion phase
are $\sim$$3\,\mathrm{cm}$ ($\sim$$10^{-22}$ at
$10\,\mathrm{kpc}$). This is comparable to, but somewhat larger than
what \cite{mueller:e12} found in the preexplosion phase of their
models. This may be due the different progenitor models used and/or to
the rather large inner boundary radius of their models in the
preexplosion phase. Our typical $|h|$ are also quantitatively
consistent with the findings of the simpler 3D simulations of
\cite{scheidegger:10} and \cite{kotake:09,kotake:11}, but are a factor
of a few smaller than predictions from 2D simulations (e.g.,
\citealt{marek:09b,yakunin:10,murphy:09}).

Figure~\ref{fig:gwspect} contrasts the angle-averaged characteristic
GW strain spectra $h_\mathrm{char}(f)$ \citep{flanhughes:98} of our
models with the broadband design noise levels of advanced-generation
GW interferometers, assuming a source distance of $10\,\mathrm{kpc}$.
The spectra are scaled with a factor of $f^{-1/2}$ to allow one-to-one
comparison with the detector one-sided amplitude spectral noise
density $\sqrt{S(f)}$, which has units of $\mathrm{Hz}^{1/2}$. Most of
the detectable emission is within $\sim$$60-1000\,\mathrm{Hz}$ and at
essentially the same level of
$\sim$$2-6\times10^{-23}\,\mathrm{Hz}^{-1/2}$. A galactic event (at
$10\,\mathrm{kpc}$) appears to be well detectable by the upcoming
generation of detectors. All four models, while having distinct
individual $h_+$ and $h_\times$ time series that vary greatly in the
time domain, exhibit essentially the same robust spectral features,
independent of $f_\mathrm{heat}$ and the exact postbounce time the
individual models are evolved to. The low-frequency to
intermediate-frequency emission is most likely due to prompt
convection in the early postbounce phase, while the high-frequency
peaks at $\sim$$400\,\mathrm{Hz}$ and $\sim$$900\,\mathrm{Hz}$ are
most likely due to the deceleration of downflows at the protoneutron
star surface and protoneutron star convection. A more detailed
investigation of these features must be left to future work, since it
would require multiple quadrupole integrals to isolate emission
regions as done, e.g., by \cite{murphy:09}. Our present simulations
provide only one global quadrupole integral and we do not have
sufficiently finely sampled output for a postprocessing analysis of the
GW signal.

The total energy emitted in GWs is $3-4 \times 10^{-10}\,M_\odot\,c^2$
in all our models and about $50\%$ of the emitted energy is due to the
higher-frequency GW emission at later postbounce times. This finding
is consistent with the 3D results of
\cite{scheidegger:10}. \cite{mueller:e12}, on the other hand, found
emitted GW energies of only $\sim$$10^{-11}\,M_\odot c^2$. Their models
do not include prompt convection and emit most of their GW energy at
frequencies below $\sim$$400-600\,\mathrm{Hz}$. This, again, may be
due to the different considered progenitor structures and/or to the
inner boundary of their simulations.

\section{Discussion and Conclusions}
\label{sec:conclusions}

We have carried out four 3D general-relativistic core collapse and
postbounce simulations of the $27$-$M_\odot$ solar-metallicity
progenitor of \cite{whw:02}, systematically varying the rate of
neutrino energy deposition to study the effect of variations in
neutrino heating on the 3D postbounce evolution in general and on the
standing accretion shock instability (SASI) in particular.  These
simulations neither employed an artificial inner boundary nor did they
make any symmetry assumptions or approximations for the gravitational
field. The resolution of our simulations is nearly twice as high and
we carried them out for nearly twice as long as the only previous 3D
GR study of \cite{kuroda:12}.

For neutrinos, we used an energy-averaged (gray) three-species
neutrino leakage/heating scheme in the postbounce phase, whose only
free parameter is a scaling factor in the energy deposition rate. The
leakage scheme captures the essential aspects of neutrino cooling,
lepton number exchange, and neutrino heating as predicted by fully
self-consistent 1D and 2D neutrino radiation-hydrodynamics
simulations.  Importantly, our simulations do not suffer from the
limitations of simpler analytic ``light-bulb'' heating/cooling
schemes, which cannot capture the contraction and deleptonization of
the protoneutron star and result in artificially large shock radii and
overestimated advection times through the postshock region
(\citealt{richers:13}, \emph{in prep.}).  The light-bulb
approach, due to its simplicity and low computational cost, is being
employed in many contemporary 3D simulations (e.g.,
\citealt{nordhaus:10,hanke:12,burrows:12,murphy:12,dolence:12}).
However, as pointed out by \cite{hanke:12} and \cite{mueller:12b},
light-bulb calculations may yield qualitatively incorrect results for
the postbounce hydrodynamics and the respective roles and relevance of
neutrino-driven convection and the SASI.

Our approach was designed specifically to avoid the problems of the
light-bulb scheme and provide a realistic postbounce setting for more
robust conclusions on the postbounce evolution and the role of
hydrodynamic instabilities. At the same time, our leakage/scheme is
still computationally much cheaper and simpler than the approximate
gray or energy-dependent 3D neutrino transport schemes of
\cite{kuroda:12}, \cite{takiwaki:12}, and
\cite{mueller:e12,wongwathanarat:10}. This affords us with the ability
to carry out parameter studies with high numerical resolution as
presented in this work for the $27$-$M_\odot$ progenitor.

\cite{mueller:12b} previously carried out an axisymmetric (2D)
simulation of the same $27$-$M_\odot$ progenitor with their 2D GR
radiation-hydrodynamics code. They found neutrino-driven convection to
be suppressed due to the high postbounce accretion rate and, thus,
short advection time through the convectively unstable gain layer. The
SASI is the primary instability in their simulation and seeds
convection, which grows only as a secondary instability once the SASI
has reached non-linear amplitudes.

Our models show instead early and strong growth of convective
instability. It is initially prompt, driven by the negative entropy
gradient left behind by the stalling shock. Subsequently, convection
is driven by neutrino energy deposition in the gain
layer. Neutrino-driven convection first manifests itself in
small-scale local rising hotter and sinking cooler blobs of postshock
material. In models with strong neutrino heating that are trending
towards explosion, the small scale blobs combine over time to a few
large, near volume-filling high-entropy regions whose expansion pushes
out the shock. This was also observed in the high-luminosity
light-bulb simulations of \cite{burrows:12} and \cite{dolence:12}.
These large blobs lead to a low-$\ell$-mode dominated structure of the
expanding shock. The shock, however, has a complicated substructure of
protruding bumps caused by smaller-scale plumes that perturb it
locally.  Models whose shock expansion becomes dynamical, surpass the
runaway explosion criterion $\tau_\mathrm{adv} / \tau_\mathrm{heat}
\gtrsim 1$ \citep{burrows:93,janka:01} and satisfy the antesonic
condition of \cite{pejcha:12a}. Both criteria for explosion yield
predictions consistent with the trends in our models.  Interestingly,
shortly after the $\tau_\mathrm{adv} / \tau_\mathrm{heat} \gtrsim 1$
condition is met by our models, individual fluid cells behind the
shock reach positive total energy, indicating the transition to
explosion.

While neutrino-driven convection is the fastest growing and overall
dominant instability, our analysis suggests that all of our models
exhibit some growth of clearly periodic low-$\ell$ deformations of the
shock front that are characterstic of the linear phase of the SASI. As
expected from linear perturbation analysis, we find that the $\ell =
1, m=\{-1,0,1\}$ modes exhibit the fastest growth. However, our
results also show that the saturation amplitudes of the oscillatory
$\ell = 1, m=\{-1,0,1\}$ modes are, in the best case, an order of
magnitude smaller than in \cite{mueller:12b}. The SASI remains a
sub-dominant instability in all of our models. Furthermore, we find
the SASI to be strongest in the model with the least neutrino heating
and the weakest neutrino-driven convection. Models with stronger
heating and more vigorous convection have lower saturation amplitudes
of the oscillatory modes, but develop large \emph{non-oscillatory}
deformations of $\ell = 1, 2, 3$ character that are caused by low-mode
neutrino-driven convection and are unrelated to the SASI\@.

Our simulations satisfy all the requirements laid out by
\cite{mueller:12b} for the development of strong SASI in the
$27$-$M_\odot$ progenitor: GR gravity, an EOS that results in a fairly
compact protoneutron star, and the inclusion of all neutrino species
and deleptonization of the protoneutron star. Yet, our results turn
out to be very different from what \cite{mueller:12b} found. What is
the root cause of this discrepancy? On the one hand, our simulations
are 3D, splitting, on average, the 2D $\ell = 1$ SASI power across
three azimuthal $m$ modes. This may explain lower saturation
amplitudes, but cannot explain the early growth of neutrino-driven
convection that is absent from the 2D simulation of
\cite{mueller:12b}.  On the other hand, -- and, as we are convinced,
more importantly -- our simulations used a central Cartesian
adaptive-mesh refinement (AMR) grid, which imparts perturbations of
order of $1-10\%$ onto the very early postbounce flow, seeding prompt
convection. This, in turn, acts as seed for neutrino-driven convection
in our models. The seed perturbations are sufficiently large for
convection to develop despite the high accretion rate and
correspondingly short advection time through the gain
layer. Neutrino-driven convection becomes dominant and limits the
growth of the SASI, in agreement with the 2D work of \cite{scheck:08}.
We expect any 3D simulation relying on 3D Cartesian AMR with similar
resolution to have similarly large seed perturbations for
convection. The recent 3D light-bulb simulations of
\cite{burrows:12,murphy:12,dolence:12} are all subject to these
perturbations.

The question of the magnitude of seed perturbations was not raised by
\cite{mueller:12b}, who used a spherical-polar grid that leads to only
minute perturbations from the growth of numerical noise during
collapse. Is the almost perfectly spherical postbounce state of
\cite{mueller:12b} representative of nature or should one expect
significant asphericities to be present in the outer core?  Some
guidance on the size of perturbations induced by turbulent convection
during late time burning in core-collapse supernova progenitors is
already available from the 2D and 3D simulations of Meakin, Arnett,
and collaborators \citep{meakin:06,meakin:06b,meakin:07b,arnett:11a}.

There are two important results from these multi-dimensional stellar
evolution calculations that pertain to the expected density
perturbation amplitudes in precollapse cores.  First, 2D and 3D
simulations of the  oxygen shell burning dominated phase in a 23-\msun star 
\citep{meakin:07b} have clarified the basic mechanism responsible for
the origin of the fluctuations. In short, \cite{meakin:07a} found
that the root-mean-square (rms) density fluctuations are largest at the
convective boundaries.  By interpreting the dynamics of the convective
boundary layer in terms of g-modes excited by the turbulent
convection, it was shown that the rms density fluctuation
amplitude can be related directly to the background stellar structure
and the Mach number of the convective flow, with

\begin{align}
\frac{\delta \rho}{\rho} \sim  M_c^2 + \frac{v_s \omega_{\mathrm{BV}} M_c}{g}\,\,,
\end{align}

\noindent where $M_c$ is the rms Mach number of the convective flow,
$\omega_\mathrm{BV}$ is the Brunt-V\"ais\"al\"a frequency in the
stable layer adjacent to the convection zone, $v_s$ is the sound
speed of the gas, and $g$ is the gravitational acceleration.  The
first term on the right hand side, which is very small, is relevant to
the interior of the convection zone, where density fluctuations arise
solely from the presence of velocity fluctuations in a nearly
adiabatic layer.  The second term is significantly larger and applies
to the stable layers bounding the convection zone, reflecting the
excitation of fluid motions in these regions in the form of internal
waves (predominantly g-modes).

The $27$-$M_\odot$ progenitor of \citealt{whw:02} has a turbulent Mach
number of $\sim$0.1 to 0.2 in the silicon burning convective shell
overlying the core, and two peaks in $\omega_\mathrm{BV}$ of
importance: the peak corresponding to the inner edge of the active
silicon burning shell (corresponding to the outer edge of the iron
core), and a peak deeper in associated with the outermost extent of
the now extinguished silicon burning core.  Both peaks have values of
$ v_s \omega_\mathrm{BV}/g$ of $\sim$1, indicating that rms density
fluctuations at these locations will be of order the turbulence Mach
number of the convection, or $\sim10-20\%$. The spike in
$\omega_\mathrm{BV}$ associated with the outer extent of the silicon
core burning epoch will be accreted into the shock within
$\sim$$15\,\mathrm{ms}$ of bounce, while the edge of the iron core
will be accreted a little later, at $\sim$60~ms after bounce.
  
The second result from the multi-D stellar convection simulations of
Meakin and Arnett involves the interaction of nuclear burning shells
at late times. While the results on boundary layer fluctuations
described above are considered to be robust by those authors, the
presence of two or more convective shells in close proximity, as found
in late burnings stages, has been found to drive additional motion at
the convective boundaries and correspondingly larger density
fluctuation amplitudes.  In the most relevant case of a silicon
burning shell around an iron core, the interaction between the
silicon, oxygen, neon, and carbon shells were found to produce a
dramatic increase in boundary layer distortion, eventually leading to
a complete disruption and mixing of the multi-shell burning region
\citep{meakin:06,arnett:11a}.  This result is likely to be due, at
least in part, to the inconsistency between the initial stellar model
used (based on mixing length theory) and a more realistic turbulent
convection as represented by the numerical simulation. Judging the
robustness of these shell-interaction results, however, awaits 3D
simulations since all of the multi-shell calculations performed to
date have been restricted to 2D geometry which is known to result in
exaggerated velocities in regions of thermal convection.  From this
body of work, it would appear that the presence of density
fluctuations with amplitudes of at least 1\%, and possibly as large as
10 to 20\%, should be expected in the material accreting into the
shock at early postbounce times in a collapsing iron core.

The fast growth of neutrino-driven convection in our current models is
almost certainly caused by the large seed perturbations from our
Cartesian AMR grid. In 2D simulations, the growth of neutrino-driven
convection may go along with SASI growth or, if not genuine SASI, then
at least large-scale \emph{oscillatory} low-$\ell$ deformations of the
shock front
\citep{mueller:12b,burrows:12,fernandez:09b,scheck:08}. Our 3D models
do not exhibit any large-scale oscillatory features. Rather, models
evolving towards an explosion develop non-oscillatory large-scale
asphericities at late times and produce a globally aspherical
explosion morphology without a need for SASI-driven $\ell = 1$
deformations. This qualitative finding is in agreement with the
results of the convection-dominated 3D Newtonian light-bulb
calculations of \cite{burrows:12} and \cite{dolence:12}.  The
late-time development of SASI-like oscillatory behavior seen in 2D
simulations that are initially convection dominated (e.g.,
\citealt{marek:09,mueller:12a}) may thus be an artifact of
axisymmetry, but further work is required to solidify this conclusion.

The next galactic core-collapse supernova will reveal its inner
workings by means of its neutrino and gravitational-wave (GW)
signals. Both will provide key insight into the thermodynamics and
multi-D dynamics of the protoneutron star and the postshock region
(e.g., \citealt{ott:09,lund:10,lund:12,oconnor:13}). While our
neutrino treatment is too simplistic to yield quantitatively
interesting predictions of the neutrino signal, we are in a good
position to study the GW emission from accelerated quadrupole mass
motions in our models: For the first time, we have extracted GWs from
full 3D GR collapse and postbounce core-collapse supernova
simulations. We find a strong burst of GWs associated with
early-postbounce prompt convection with frequencies around
$\sim$$100-200\,\mathrm{Hz}$, a subsequent almost quiescent phase,
followed by higher-frequency ($400-1000\,\mathrm{Hz}$) emission, whose
amplitudes are dominated by the deceleration of undershooting
convective plumes at the edge of the protoneutron star
(cf.~\citealt{murphy:09}).  If convection (prompt and/or
neutrino-driven) does not develop early, the GW signal would not have
a strong initial burst, but rather a slow rise to smaller amplitudes
at later times, when the SASI becomes strong.  This is a key
difference and may allow GW data analysts to distinguish between
convection-dominated and SASI-dominated postbounce evolution in the
next galactic core-collaspe supernova.  The design sensitivities of
advanced-generation GW detectors such as Advanced LIGO, Advanced
Virgo, or KAGRA are likely to be sufficient to detect the collapse and
neutrino-driven explosion in our $27$-$M_\odot$ progenitor throughout
the Milky Way.  While different in detail, our results for the GW
signature are generally consistent with what was found for other
progenitors in the 2D first-principles simulations of \cite{marek:09b}
and \cite{yakunin:10}. Our GW signals have higher amplitudes and
characteristic frequencies than predicted by the 3D simulations of
\cite{mueller:e12}, who employed an artifical inner boundary that was
moved in according to an analytic prescription.

There are a number of shortcomings and limitations of the simulations
presented here that must be mentioned and can be removed only by
future work.  As is well known and has been pointed out recently by
\cite{hanke:12} in the core collapse context, in 3D, turbulent power
cascades to small scales. Low resolution in 3D may artificially keep
power at large scales and may thus lead to an overestimate of the
positive effect of neutrino-driven convection.  While our effective
angular and radial resolution in the postshock gain layer is
comparable to the highest resolution considered by \cite{hanke:12}, we
agree with their assessment that understanding the resolution
dependence of 3D results is of great importance.  We will carry out a
resolution study in future work.

The second major limitation of our simulations is our Cartesian AMR
grid and the fact that we must let the nascent supernova shock pass
two mesh refinement boundaries before tracking its further evolution
by AMR. This induces large perturbations leading to the growth of
prompt and neutrino-driven convection in all of our models. These
large and essentially unavoidable seed perturbations for prompt and
neutrino-driven convection make it difficult to draw conclusions on
which hydrodynamic instability dominates in the early postbounce
phase. This limitation is shared by other Cartesian AMR schemes. It
could possibly be avoided in future work by extending our
spherical-polar grid blocks all the way into the protoneutron star
core and using a single high-resolution Cartesian mesh only in the
innermost few kilometers. Also in future work, we intend to carry out
a study in which we map a postbounce profile from a 1D collapse
simulation onto our 3D grid after the shock has passed the radii of
the inner refinement levels. This should allow us to investigate the
role of seed perturbations in a more controlled way.

A third major limitation of our work is the reliance on our simple
gray heating/leakage scheme. While superior to the light-bulb
approach, it cannot replace the energy-dependent neutrino
radiation-hydrodynamics treatment that has proven to be crucial for
reliable conclusions on the neutrino mechanism (e.g.,
\citealt{mueller:12a} and references therein). The set of 3D
general-relativistic hydrodynamics simulations presented here required
about $\sim$$20$ million CPU hours to complete. Adding
energy-dependent 3D neutrino transport will increase the computational
complexity by an order of magnitude. Novel, highly efficient and
scalable approaches to 3D neutrino transport will be needed to address
this problem
\citep{sumiyoshi:12,abdikamalov:12,zhang:13,radice:12rad}.

\section*{Acknowledgement}
We acknowledge helpful discussions with Dave Arnett, Adam Burrows,
Sean Couch, Luc Dessart, Thierry Foglizzo, Uschi~C.~T.~Gamma, Sarah
Gossan, Raph Hix, H.-Thomas Janka, Peter Kalmus, Hannah Klion, Io
Kleiser, Jim Lattimer, Bernhard M\"uller, Jeremiah Murphy, David
Radice, Luke Roberts, Jason Nordhaus, Ken Nomoto, Jerome Novak, Tony
Piro, Sherwood Richers, and members of our Simulating eXtreme
Spacetimes (SXS) collaboration (http://www.black-holes.org). This
research is partially supported by NSF grant nos.\ AST-0855535,
AST-1212170, PHY-0904015, PHY-1151197, OCI-0905046, and OCI-0941653,
by the Sloan Research Foundation, and by the Sherman Fairchild
Foundation.  CR acknowledges support by NASA through Einstein
Postdoctoral Fellowship grant number PF2-130099 awarded by the Chandra
X-ray center, which is operated by the Smithsonian Astrophysical
Observatory for NASA under contract NAS8-03060.  RH acknowledges
support by the Natural Sciences and Engineering Council of Canada.
The simulations were performed on the Caltech compute cluster
``Zwicky'' (NSF MRI award No.\ PHY-0960291), on supercomputers of the
NSF XSEDE network under computer time allocation TG-PHY100033, on
machines of the Louisiana Optical Network Initiative under grant
loni\_numrel07, and at the National Energy Research Scientific
Computing Center (NERSC), which is supported by the Office of Science
of the US Department of Energy under contract DE-AC02-05CH11231. The
multi-dimensional visualizations were generated with the open-source
\code{VisIt} visualization package
(\url{https://wci.llnl.gov/codes/visit/}). All other figures were
generated with the \code{Python}-based \code{matplotlib} package
(\url{http://matplotlib.org/}).

\end{document}